\documentclass{osa-article}

\journal{oe}
\articletype{Research Article}
\begin{document}
\newcommand{\fft}[1]{\widetilde{#1}}
\newcommand{\ifft}[1]{\widehat{#1}}

\title{Numerical study of solitonic pulse generation in the self-injection locking regime at normal and anomalous group velocity dispersion}

\author{Nikita~M.~ Kondratiev,\authormark{1+} Valery~E.~ Lobanov,\authormark{1*} Evgeny~A.~ Lonshakov,\authormark{1} Nikita~Yu.~ Dmitriev,\authormark{1,2} Andrey~S.~ Voloshin,\authormark{1} and  Igor~A.~Bilenko\authormark{1,3} }

\address{\authormark{1}Russian Quantum Center, 143026, Skolkovo, Russia\\
	\authormark{2}Moscow Institute of Physics and Technology, 141700, Dolgoprudny, Russia\\
	\authormark{3}Faculty of Physics, Lomonosov Moscow State University, 119991, Moscow, Russia}

\email{\authormark{*}v.lobanov@rqc.ru}
\email{\authormark{+}noxobar@mail.ru}



\begin{abstract}
We developed an original model describing the process of the frequency comb generation in the self-injection locking regime and performed numerical simulation of this process. Generation of the dissipative Kerr solitons in the self-injection locking regime at anomalous group velocity dispersion was studied numerically. Different regimes of the soliton excitation depending on the locking phase, backscattering parameter and pump power were identified. It was also proposed and confirmed numerically that self-injection locking may provide an easy way for the generation of the frequency combs at normal group velocity dispersion. Generation of platicons was demonstrated and studied in detail. The parameter range providing platicon excitation was found. 
\end{abstract}


%\maketitle

\section{INTRODUCTION}
The effect of the self-injection locking (SIL) is well-known for many years in the theory of oscillations, radiophysics and optics and is actively used for the stabilization and spectral purification of the corresponding generators \cite{Ohta1,Chang2,magnetron1,magnetron2,gyrotron2,7119883}. In optics it allows to obtain sub-kHz or even sub-Hz generation linewidth using compact semiconductor lasers locked to high-Q optical microresonators \cite{Patzak1983,Agrawal1984,Oraevsky2001,Liang2010,Liang2015,Kondratiev:17,PhysRevApplied.14.014036,jin2020hertzlinewidth}, e.g. whispering-gallery-mode (WGM) microresonators \cite{BRAGINSKY1989393,1588878,Lin2017}. Last years it has attracted even more attention due to the possibility of using such stabilized lasers as pump sources for the realization of the nonlinear processes in the same microresonators, simultaneously used for laser linewidth reduction. In particular, the generation of the mode-locked microresonator-based frequency combs in the form of the dissipative Kerr solitons (DKS) \cite{herr2014temporal, Kippenbergeaan8083} was demonstrated in bulk and on-chip microresonators pumped by the commercial diode lasers \cite{Pavlov2018,Raja2019, Shen2020, jin2020hertzlinewidth, Voloshin2020}. This paved the way for the creation of compact and affordable sources of the low-noise frequency combs, which are in great demand for spectroscopy, metrology, telecommunications, etc \cite{Suh600,9195660, Marin-Palomo2017,Suh2019,Kippenbergeaan8083}. Moreover, such method of the soliton excitation allows to avoid some difficulties practically inevitable at soliton generation with free-running laser. The obstacle lies in the temperature drop the resonator experiences when the pump laser transits from the effectively blue-detuned (high intracavity power) to the red-detuned (lower intracavity power) state \cite{herr2014temporal}. This jump can lead to a shift in the operating point and disruption of generation. In the self-injection locking regime, when pump frequency is locked to the microresonator eigenfrequency, this problem can be overcome easily. Also, it was demonstrated that dynamics of the soliton generation in the self-injection locking regime differs significantly from the dynamics of the same process in the unlocked regime \cite{Voloshin2020}. However, despite the wonderful experimental results the comprehensive theory of the process of the frequency comb generation in the self-injection locking regime has not been developed yet. Moreover, existing linear theories of the self-injection locking can not predict soliton generation because {it is impossible to obtain enough value of the pump frequency detuning in the linear regime}. The development of such theory becomes even more important, since recently generation of the solitonic pulses at normal group velocity dispersion (GVD) has been demonstrated in the self-injection locking regime \cite{ Lihachev2020cleo,jin2020hertzlinewidth}. Such approach allows to simplify the setup since it does not require additional pump modulation or bi-chromatic pump \cite{Lobanov_2015,Liu:17,Lobanov2019}  or complex systems [e.g. two coupled resonators] necessary for the realization of the controllable mode interactions \cite{Liu:14,Xue2015,Lobanov2015,Lobanov2017, Kim:s, doi:10.1002/lpor.201500107}. {Generation of the dark solitons or platicons which can be considered as a flat-top pulse between two dark solitons was found to be more efficient in terms of the pump-to-comb conversion efficiency than generation of bright solitons \cite{ doi:10.1002/lpor.201600276, Kim:s, Jang:s}.} %In many cases, it more convenient to study the dynamics of pulses (platicons) than the dynamics of dips (dark solitons).}
In our work we develop an original model describing the process of the frequency comb generation in the self-injection locking regime, combining approaches described in \cite{Voloshin2020} and \cite{Kondratiev2019}, and perform numerical simulation of this process for both anomalous and normal GVD. We show that nonlinear frequency shift allows to reach the desired pump frequency detuning providing DKS generation inside the locking band. For the anomalous GVD different regimes of the dissipative Kerr soliton generation are identified in the locked and unlocked state. These regimes are found to depend on the laser-microresonator phase distance [locking phase], backscattering parameter and pump power.  For the normal GVD different regimes of the frequency comb generation, including generation of platicons, are demonstrated numerically. The values of the pump amplitude and backscattering coefficient allowing for the platicon excitation were found.

\section{Complete model}

%\begin{figure}[ht]
%\centering
%\includegraphics[width=0.7\linewidth]{Figs/Scheme.pdf}
%\caption{Schematic description of the self-injection locking of the laser diode. }
%\label{fig:scheme}
%\end{figure}

%We consider the model of the self-injection locking shown in Figure \ref{fig:scheme}. 
%We begin with the equations, obtained for the SIL effect in our previous work \cite{Kondratiev:17}, and
For numerical analysis we developed a combined model describing simultaneously pump laser dynamics and field evolution inside microresonator.
The laser field is considered to be single-mode and normalized to the photon concentration $\vec E_{\rm L}=\sqrt{\frac{2\hbar\Omega_l}{\epsilon_0n_L^2}}\vec e_l(\vec r) A_l e^{-i\Omega_l t}$, $n_L$ is the laser refraction index, $\Omega_l$ and $\vec e_l$ is the laser mode frequency and spatial form. 
To construct the SIL theory including comb generation effect we expand the microresonator field into sum of the modal oscillations similar to \cite{Kondratiev2019}: 
$\vec E_{\rm W}=\sqrt{\frac{2\hbar\Omega_l}{\epsilon_0n_L^2}}\sum (\vec e_\mu^+(\vec r)A_\mu^+ +\vec e_\mu^-(\vec r)A_\mu^- )e^{-i\omega_\mu^{(1)}t}$. Here we define the modal amplitudes' main frequencies to be on the FSR-grid $\omega_\mu^{(1)}=\omega_0+\mu D_{1_W}$, where $\omega_0$ is the microresonator mode eigenfrequency nearest to $\Omega_l$ and $D_1$ is the first-order dispersion coefficient [intermode distance or free spectral range -- FSR -- of modes near $\omega_0$]. The $\vec e^{\pm}_\mu$ are the spatial profiles of the modes.
The field amplitudes are normalized to photon concentration using the same laser-referred coefficient to simplify the expressions. 
We modify accordingly the equation system for the SIL effect from \cite{Kondratiev:17}, thus combining it with the results of \cite{Kondratiev2019} for the high-finesse microresonator and obtain the coupled mode equation system (CMES) \cite{PhysRevA.82.033801} in the form
\begin{align}
\label{Model_carr}
\frac{dN}{d\tau} =& J_N - \frac{\kappa_N}{\kappa_0} N - N g_l |A_l|^2,\\
\label{Model_las}
\frac{dA_l}{d\tau}=&\left(-i\xi_0-iv_\xi \tau+(1+i\alpha_g)Ng_l-\frac{\kappa_l}{\kappa_0}\right)A_l-e^{i\Omega_l t}\sum_\mu \tilde\kappa_{\rm Laser}A^-_\mu e^{-i\omega_\mu^{(1)} (t-t_s)},\\
\label{Model_forw}
\frac{dA^+_\mu}{d\tau}=&\frac{1}{\kappa_0}\left(-\kappa_\mu-i2(\omega_\mu-\omega_\mu^{(1)})\right)A^+_\mu +i\beta_\mu A^-_\mu +i\tilde g_\mu S^+_\mu -\tilde \kappa_{\rm WGR}e^{i\omega_\mu^{(1)} t}\delta_{0\mu}A_le^{-i\Omega_l (t-t_s)},\\
\label{Model_back}
\frac{dA^-_\mu }{d\tau}=&\frac{1}{\kappa_0}\left(-\kappa_\mu-i2(\omega_\mu-\omega_\mu^{(1)})\right)A^-_\mu +i\beta_\mu A^+_\mu +i\tilde g_\mu S^-_\mu.
\end{align}
{The obtained equations are similar to those in \cite{Bao2019}, but the laser is assumed single-mode here and the coupling is realised by via the backward wave.}
The equation \eqref{Model_carr} describes the carrier concentration dynamics and \eqref{Model_las} -- field amplitude in the laser. The last term of \eqref{Model_las} is the sum of the fields coming from the WGM cavity (backward wave). In practice the sum in \eqref{Model_las} is calculated so that $\left|\Omega_l-\omega_\mu^{(1)}-{\rm Im}\left[\frac{dA_\mu^-/dt}{A_\mu^-}\right]\right|<5\kappa_l$, where $\kappa_l$ is the laser cavity linewidth, to avoid modeling of redundant fast oscillating terms.
The second pair \eqref{Model_forw}-\eqref{Model_back} describes the WGM field in the high-finesse limit \cite{Kondratiev2019} [note the $\delta$-symbol in the  pump term of \eqref{Model_forw}]. The terms with $\beta$ stand for the forward-backward mode coupling. The backward wave \eqref{Model_back} is excited only through this term. The forward wave has two pumps: the backward wave ($i\beta_\mu A^-_\mu$ term) and the laser (the last term). 
Here $N$ is the carrier concentration,
$J_N=\frac{2I}{eV_l\kappa_0}$ is the normalized injection current ($I$ is the current, $e$ is elemental electron charge and $V_l$ is the laser mode volume), 
$\kappa_N$, $\kappa_l$ and $\kappa_\mu$ are the relaxation rates of the inverse population, laser mode and $\mu$-th microresonator mode,
$\tau=\kappa_0 t/2$ is the normalized time,
$t_s$ is the one-way-trip-time from the laser to the microresonator defining the locking phase \cite{Kondratiev:17},
$g_l$ is the normalized linear laser gain,
$\alpha_g$ is the Henry factor,
$\tilde g_\mu$ is the normalized nonlinear coefficient. The microresonator modes are numerated from the nearest to the laser cavity frequency $\Omega_l$, 
%the $\omega_\mu^{(1)}=\omega_0+\mu D_{1_W}$ is the first order frequency dispersion (FSR-grid) of the microresonator with $\omega_0$ being its central frequency,
{so that we can assume $\Omega_l=\omega_0$ and $\xi_0=2(\Omega_{\rm init}-\omega_0)/\kappa_0$ is the laser cavity detuning.} The $v_\xi$ is the tuning speed,
%$\Delta\Omega_{\mu l}=\omega_\mu^{(1)}-\Omega_l$,
$\beta_\mu$ is the normalized forward-backward mode coupling or backscattering coefficient for the $\mu$-th mode [equal to the mode splitting in units of $\kappa_0$]. 
{The $\omega_\mu-\omega_\mu^{(1)}$ is the dispersion term [real WGM mode frequencies are $\omega_\mu=\omega_\mu^{(1)}+D_2\mu^2+D_3\mu^3+...$].}
The nonlinear sums, representing Kerr effect, are\cite{Kondratiev2019}
\begin{align}
S^+_\mu=\ifft{\fft{A^+}\fft{A^+}\fft{A^{+*}}}_\mu+2\Theta' A^+_\mu\sum|A^-_k|^2,\\
S^-_\mu=N^2\fft{\ifft{A^-}\ifft{A^-}\ifft{A^{-*}}}_\mu+2\Theta'A^-_\mu\sum|A^+_k|^2,
\end{align}
where $\widetilde x_m=\sum_{\nu=0}^{M-1}x_\nu e^{-2\pi i\nu m/M}$ is the discrete Fourier transform and  $\widehat x_\nu=\frac{1}{M}\sum_{p=0}^{M-1}x_p e^{2\pi i\nu p/M}$ is the inverse one, where $M$ is the number of modes. The $\Theta'$ is a cross-interaction coefficient that is assumed to be unity here as for modes of the same polarization.
The initial detuning $\xi_{0}$ can be taken arbitrary.% It is convenient to shift the initial frequency $\xi_{0}$ over $\alpha_g\frac{\kappa_l}{\kappa_0}$ to have the tuning curve going through (0,0) point.

The generalized coupling coefficients $\tilde\kappa_{\rm Laser}$ and $\tilde\kappa_{\rm WGR}$ can be expressed in terms of the 
WGM coupling efficiency $\eta_\mu=\frac{\kappa_c}{\kappa_\mu}$ ($\kappa_c$ is mode coupling rate \cite{Gorodetsky:00}),
amplitude reflectivity and transmission of the laser output mirror $R_o$ and $T_o$,
WGM and laser normalized roundtrip times $\tau_W$ and $\tau_L$ [note that these times 
%$\tau_W=\frac{c\kappa_0}{4\pi R n_{\rm eff}}$
are in general different from the inverse finesse $\mathcal{F}\approx\frac{D_1}{\kappa_0}$ 
as the former is determined by the effective refraction index and the latter by the group index]:
\begin{align}
\tilde\kappa_{\rm Laser}=&\frac{T_o}{\tau_L}\sqrt{\frac{2\eta_\mu\kappa_\mu\tau_W}{\kappa_0}}\sqrt{\frac{n_{\rm c}S_{\rm c}}{n_{\rm L}S_{\rm Laser}}},\\
\tilde\kappa_{\rm WGR}=&\sqrt{\frac{2\eta_\mu\kappa_\mu}{\kappa_0\tau_W}}\frac{T_o}{R_o}\sqrt{\frac{n_{\rm L}S_{\rm Laser}}{n_{\rm c}S_{\rm c}}}.
\end{align}
Here for WGM microresonator we have the multiplier $\frac{1}{R_o}$ to take the direction of laser power into account. 
The last terms with the coupler and laser mode areas and refraction index ratio is to convert the field amplitudes preserving the power flow. 

The Table \ref{tab:Param} provides the main parameters of the system and the values used in the modeling. {We use the parameters, common for the real-world microresonators, used in the experiments.  While the developed model allows to take high-order dispersion terms into account, in this study we limit our study to the second order dispersion, so that $\omega_\mu-\omega_\mu^{(1)}=d_2\mu^2$ and $d_2=2D_2/\kappa_0$ is the normalized GVD coefficient.} %The system already has a lot of free parameters and the high order dispersion should be a separate in-depth study.}
{Another free parameter is the tuning speed $v_\xi$. In unlocked regime it should also be great enough to pass the transient chaos region \cite{Karpov2019} before all solitons die and can be used to overcome the thermal effects in experiments \cite{herr2014temporal}. In the SIL case these points looks irrelevant since the detuning is nearly fixed \cite{Voloshin2020}. The other restriction is that the tuning speed should be small enough for the transient processes to finish and for the sideband power to build up before the detuning leaves the soliton existence region. This is important for low $f$ and $\beta$; for example, for $f=1.25$ and $\beta=0.001...0.01$ transition from the cw solution to soliton generation is observed if the scan speed $v_\xi$ is reduced five times.}

%For equation \eqref{Model_forw} we also use the summation restriction $|\Delta\Omega_{\mu l}-\Im\frac{dA_l/dt}{A_l}|<100\kappa_\mu$. It is lighter than the previous to include the full locking range as the transition to locked state can occur before the turning point \cite{Kondratiev:17}.

\begin{table}[ht]
\caption{\label{tab:Param} Typical system parameters that were used for numerical simulation. We assume $\kappa_\mu=\kappa_0$, $\tilde g_\mu=\tilde g_0$ and $\beta_\mu=\beta$. The laser parameters were chosen to have critical current about 60 mA.}
\vspace{-1em}
\begin{center}
\begin{tabular}{|c|c||c|c||c|c||c|c|}
\hline
$f_e$ & (1.5;8) & $\alpha_g$ & $0.0$ & $\omega_{0}$  & $2\pi\times220$ THz& $\beta$ & (0.03;0.5) \\\hline
$\tilde\kappa_N$ & $90$ & $\tilde\kappa_l$ & $3e+5$ & $ v_{\xi}$ & $-0.1$ & $\tilde \kappa_{\rm Laser}$  & $2432$ \\\hline
$g_l$ & $7.4e-20$ m$^3$ & $\kappa_0$ & $2\pi\times1.1$ MHz & $\omega_0t_s$ & $2\pi\times(0;1)$ & $\tilde\kappa_{\rm WGR}$ & $1.72$\\\hline
$\tilde g_0$ & $4.2e-23$ m$^3$ & $d_2$ & $\pm0.04$ & $\mathcal{F}_L$ & $1.77e+04$ & $\mathcal{F}_{W}$ & $7.68e+03$\\\hline
\end{tabular}
\end{center}
\end{table}
\vspace{-1em}

\subsection{Renormalization and LLE-type equations}
{For analysis of the frequency comb generation process it is more convenient to control the normalized pump coefficient $f=\sqrt{\frac{4\tilde g_0\eta_0 P_{\rm input}}{\kappa_0 n^2\epsilon_0 V_0}}\sqrt{\frac{nS_{0}}{n_cS_{\rm c}}}$ \cite{herr2014temporal,Kondratiev2019}. This coefficient encapsulates the microresonator and coupler parameters, such as nonlinearity $\chi_3$, mode volume $V_0$, coupling coefficient $\eta_0$, mode and coupler cross-section and refractive index ratios (which is usually assumed to be unity). This coefficient has quite natural scale of the nonlinearity threshold: $f=1$ roughly corresponds to the first sideband generation and resonance curve bistability appearance. The results obtained with this parameter are universal and easily recalculated for each particular system.}
Using stationary solutions of \eqref{Model_carr}-\eqref{Model_las} with zero feedback we can estimate the laser power and get the expected normalized pump as following:
\begin{align}
f_e=&\sqrt{\tilde g_0}\tilde \kappa_{\rm WGR}\sqrt{\frac{g_l J_N-\tilde\kappa_N\tilde\kappa_l}{g_l\tilde\kappa_l}},
\end{align}
where we introduce normalized losses $\tilde\kappa_{N,l}=\tilde\kappa_{N,l}/\kappa_0$ for convenience.
The simulations show that the actual pump coefficient $f=\sqrt{\tilde g_\mu}\tilde \kappa_{\rm WGR}|A_l|$ quickly tends to this value $f\rightarrow f_e$ and do not change significantly. Introducing the threshold current $J_{\rm th}=\frac{\tilde\kappa_N\tilde\kappa_l}{g_l}$ for the diode emission start and the critical current at which the nonlinearity manifests itself $J_{\rm cr}=\frac{\tilde\kappa_l}{\tilde g_0\tilde \kappa_{\rm WGR}^2}+J_{\rm th}$ (it corresponds to $f=1$), we can write $J_N=(J_{\rm cr}-J_{\rm th})f_e^2+J_{\rm th}$. Now we can perform the final renormalization of the system to show the resemblance with the former results. 
Substituting $A^\pm=a^\pm/\sqrt{\tilde g_0}$ and $A_l=\sqrt{\frac{g_l J_N-\tilde\kappa_N\tilde\kappa_l}{g_l\tilde\kappa_l}}a_l=\frac{f_e}{\sqrt{\tilde g_0}\tilde\kappa_{\rm WGR}}a_l$ into Eqs.  \eqref{Model_las}-\eqref{Model_back} one can obtain the common CMES with the backward wave \cite{Kondratiev2019} with zero effective detuning and time-dependent pump
\begin{align}
\label{Model_carr_n}
\frac{dNg_l}{d\tau} =& \frac{g_l}{\tilde g_0} \frac{f_e^2}{\tilde\kappa_{\rm WGR}^2}(\tilde\kappa_l-Ng_l  |a_l|^2)+\tilde\kappa_N(\tilde\kappa_l - Ng_l),\\
\label{Model_las_n}
\frac{da_l}{d\tau}=&\left(-i\xi_0-iv_\xi \tau+(1+i\alpha_g)Ng_l-\tilde \kappa_l\right)a_l-e^{i\omega_0 t}\sum_\mu \frac{\tilde\kappa_{\rm Laser}\tilde\kappa_{\rm WGR}}{f_e}a^-_\mu e^{-i\omega_\mu^{(1)} (t-t_s)},\\
\label{Model_forw_n}
\frac{da^+_\mu}{d\tau}=&\left(-1-id_2\mu ^2\right)a^+_\mu +i\beta a^-_\mu +iS^+_\mu -f_e a_le^{i\omega_0 t_s},\\
\label{Model_back_n}
\frac{da^-_\mu }{d\tau}=&\left(-1-id_2\mu ^2\right)a^-_\mu +i\beta a^+_\mu +iS^-_\mu.
\end{align}
We also used the common assumptions $\kappa_\mu=\kappa_0$, $\beta_\mu=\beta$, $g_\mu=g_0$ and $g_l=g$. The actual pump detuning for the comb generation process, which is also the laser generation detuning $\zeta=2(\omega_{\rm gen}-\omega_0)/\kappa_0$ \cite{Kondratiev:17,Voloshin2020} is hidden inside the argument of the complex amplitude $a_l$.

{Nowadays it is quite common to use the Lugiato-Lefever-type equations (LLE) to describe soliton and frequency comb generation. To provide more insight and analogy with the known systems we perform the transformation of the CMES into the LLE. The transformation  of the microresonator part of the system to the LLE-type is straightforward \cite{PhysRevA.87.053852,Kondratiev2019}. We define the spacial fields as $a(\varphi)=\sum a_\mu^+ e^{i\mu\varphi}$ and $b(\varphi)=\sum a_\mu^- e^{-i\mu\varphi}$, multiply the equations \eqref{Model_forw_n} and \eqref{Model_back_n} with corresponding exponents and sum each up over the mode number.}
Then we consider the sum of the WGM mode amplitudes in the pump term of the laser equation \eqref{Model_las_n}. Using the $b(\varphi)$ definition, we get $\sum_\mu a^-_\mu e^{-i\omega_\mu^{(1)} (t-t_s)}=e^{-i\omega_0 (t-t_s)}\sum_\mu a^-_\mu e^{-i\mu D_1(t-t_s)}=e^{-i\omega_0 (t-t_s)} b(D_1(t-t_s))$. This reads as if the feedback amplitude was gathered from the point that is rotating around the microresonator. This has a very simple physical meaning. The symmetry of the WGM is broken with the introduction of the coupling element, providing the origin of the azimuthal angle at the touching point. This is exactly the place, where the field, going to the laser, originates. It can be shown, however, that if the CMES is written in the FSR-grid, then the corresponding LLE will be obtained for the  frame ``rotating'' with FSR angular velocity. {And that is exactly our case}. The final system is as following: 
\begin{align}
\frac{dNg_l}{d\tau} =& \frac{g_l}{\tilde g_0} \frac{f_e^2}{\tilde\kappa_{\rm WGR}^2}(\tilde\kappa_l-Ng_l  |a_l|^2)+\tilde\kappa_N(\tilde\kappa_l - Ng_l),\\
\frac{da_l}{d\tau}=&\left(-i\xi_0-iv_\xi \tau+(1+i\alpha_g)Ng_l-\tilde \kappa_l\right)a_l-\frac{\tilde\kappa_{\rm Laser}\tilde\kappa_{\rm WGR}}{f_e}e^{i\omega_0 t_s} b(D_1(t-t_s)),\\
\dot{a}=&-a+id_2\frac{\partial^2a}{\partial\varphi^2}+i\beta b(-\varphi)+ia(|a|^2+2P_b)+f_e a_le^{i\omega_0 t_s},\\ 
\label{LLEBsimp}
\dot{b}=&-b+id_2\frac{\partial^2b}{\partial\varphi^2}+i\beta^*a(-\varphi)+ib(|b|^2+2P_a),
\end{align}
where $P_a=\int |a|^2d\varphi/(2\pi)$ and $P_b=\int |b|^2d\varphi/(2\pi)$ are the averaged over the circumference intensities. This system is very similar to the one, used in \cite{Shen2020}, however the detunings, the backward wave and the feedback are treated more accurately here. We also should note that the coefficients of the laser and microresonator feedback terms are not generally equal.

We have to note that this model, like the majority of coupled-mode and LLE based models, still does not take into account some effects that can occur with wideband lasers. For example if the laser power is spectrally spread it will be harder to generate the comb. However in the SIL regime the power redistribution occurs and its squeezed line will generate the comb \cite{Pavlov2018}. In such case the locking loss will automatically indicate the loss of the soliton state, regardless of the modelling results. Another point is that after the loss of the locking the laser frequency can quickly drift away from the WGM resonance due to the thermal or any other effects.

\section{Modeling results}
The Eqs. \eqref{Model_carr}-\eqref{Model_back} were solved numerically for different combinations of significant parameters: the pump coefficient $f_e$, backscattering coefficient $\beta$, the laser-resonator delay time $t_s$ and different signs of the second order dispersion coefficient $d_2$. 
{We did not focus on taking different values of $d_2$ since, to our knowledge, it does not significantly influence the soliton and platicon dynamics, only its width. In fact it can be removed from the equations, renormalizing the azimuthal variable in LLE, and it's spectrum-scaling nature is also quite evident from the CMES. For more confidence we performed several runs with $d_2=0.01$, but did not find any differences in dynamics.}
All the parameters used are gathered into Table \ref{tab:Param}. The equations are solved by the Matlab internal implementation of an explicit Runge-Kutta (2,3) pair of Bogacki and Shampine \cite{BOGACKI1989321,Shampine}.

\begin{figure}[ht]
\centering
\includegraphics[width=0.48\linewidth]{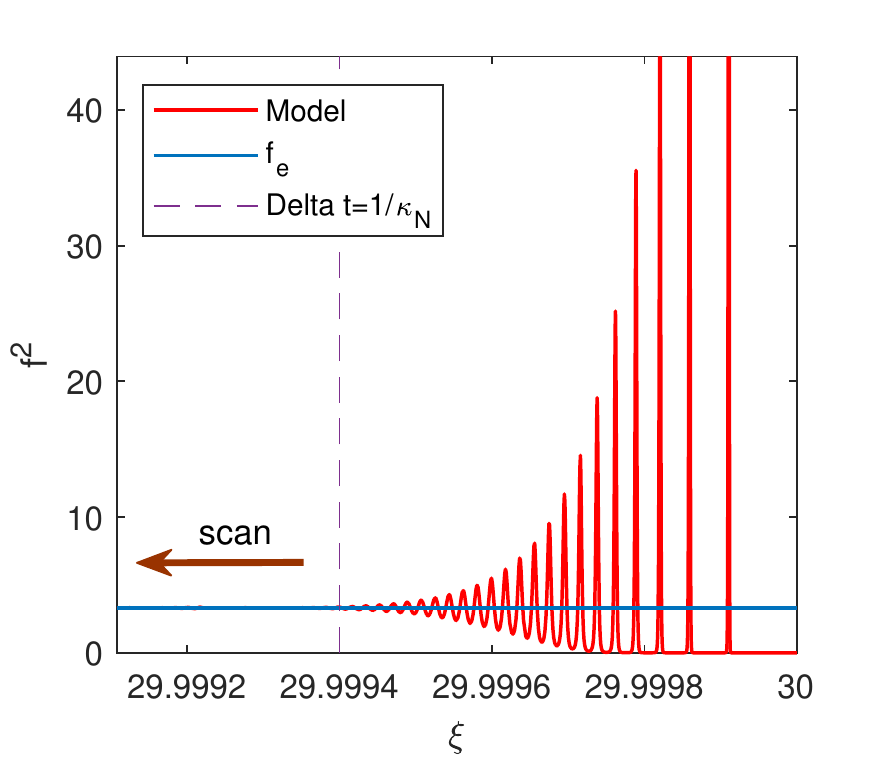}
\includegraphics[width=0.48\linewidth]{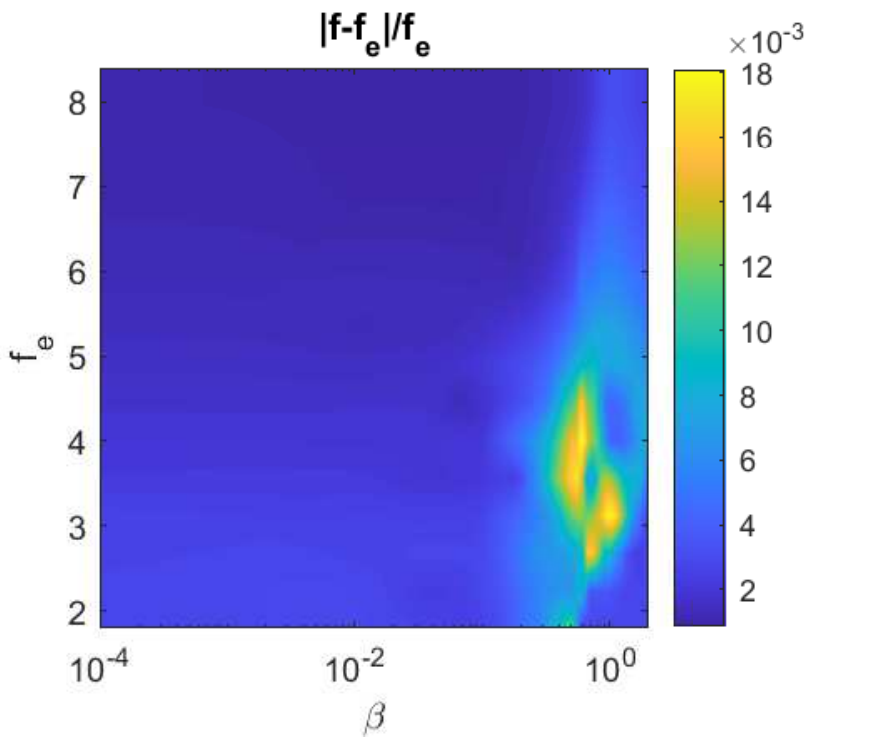}
\caption{Left: The start of the laser generation. {The actual pump amplitude $f=f_ea_l$ is shown by the red curve,} the blue horizontal line shows $f_e$ and the vertical dashed line -- the detuning after the time, referred to the charge relaxation $1/\kappa_N$. Right: average relative deviation of the actual pump amplitude $f$ from $f_e$.
}
\label{fig:Laser}
\end{figure}

The result of the simulation is the dependence of the mode amplitudes on the normalized time $\tau$. This time is then recalculated into the instantaneous laser cavity detuning $\xi=\xi_0+v_\xi \tau$. Note that in the current definitions as the nonlinear comb generation usually appears when the laser is red-detuned\cite{herr2014temporal} [here it corresponds to the negative detunings] the frequency and the detuning is to be decreased with time. So we can say that time {runs} from right to the left in the following figures.

{The main characteristic feature of the SIL is the tuning curve -- the dependence of the actual generation frequency on the laser cavity frequency or the corresponding detunings dependence $\zeta(\xi)$ \cite{Kondratiev:17,Voloshin2020}. As noted before, the laser generation detuning $\zeta=2(\omega_{\rm gen}-\omega_0)/\kappa_0$ is hidden inside the argument of the complex amplitude $A_l$ and can be naively extracted as following $\zeta_1=-\frac{d}{d\tau}\arg A_l$.} However such simple approach works only in the case of the nearly single-frequency process and give out fast oscillations if the frequency comb is generated [see, for example, light blue ``line'' in left panel in Fig. \ref{fig:f2tune}]. So the amplitude trace was divided into samples $\tau_j$-$\tau_{j+1}$ to which the Fourier transform was applied: $\tilde A_l(\tau_j,\omega)=\int_{\tau_j}^{\tau_{j+1}} A_l(\tau)e^{-i2\omega\tau/\kappa_0}d\tau$. The frequency of the highest peak was used as the instantaneous frequency $\omega_{\rm gen}$ to estimate corresponding generation detuning $\zeta_F(\tau_j)$.
To get the field profile in the microresonator, the discrete Fourier transform of the modal amplitudes should be taken over modes: $A^+(\phi)=\sum_\mu A_\mu^+e^{i\mu\phi}$.

We found that in all simulations the laser power [modulus of the field amplitude $|A_l|$)] did not exhibit significant variations, quickly reaching the stationary regime at the beginning of the simulation [see Fig. \ref{fig:Laser}, left panel]. This time was not more than the one, determined by the charge relaxation $1/\kappa_N$ time. The deviation of the actual pump amplitude $f$ from the $f_e$ was noise-like with relative amplitude no higher than 5\% [see Fig. \ref{fig:Laser}, right panel].
The only significant changes were seen only in the frequency domain ($\arg A_l$ or $\zeta$) and microresonator intracavity mode amplitudes ($A^+_\mu$ and $A^-_\mu$).

\subsection{Anomalous GVD}

We studied the dynamics of the nonlinear processes arising upon the variation of the laser cavity detuning $\xi$ in the anomalous group velocity dispersion spectral range ($d_2<0$) and different regimes were observed.
\begin{figure}[ht]
\centering
\includegraphics[width=0.48\linewidth]{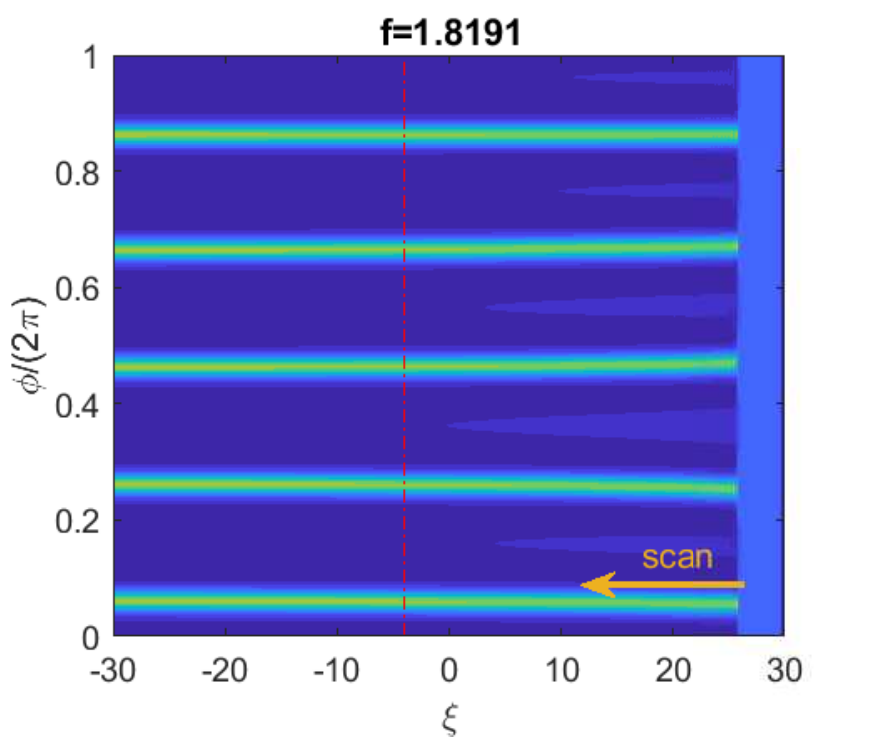}
\includegraphics[width=0.48\linewidth]{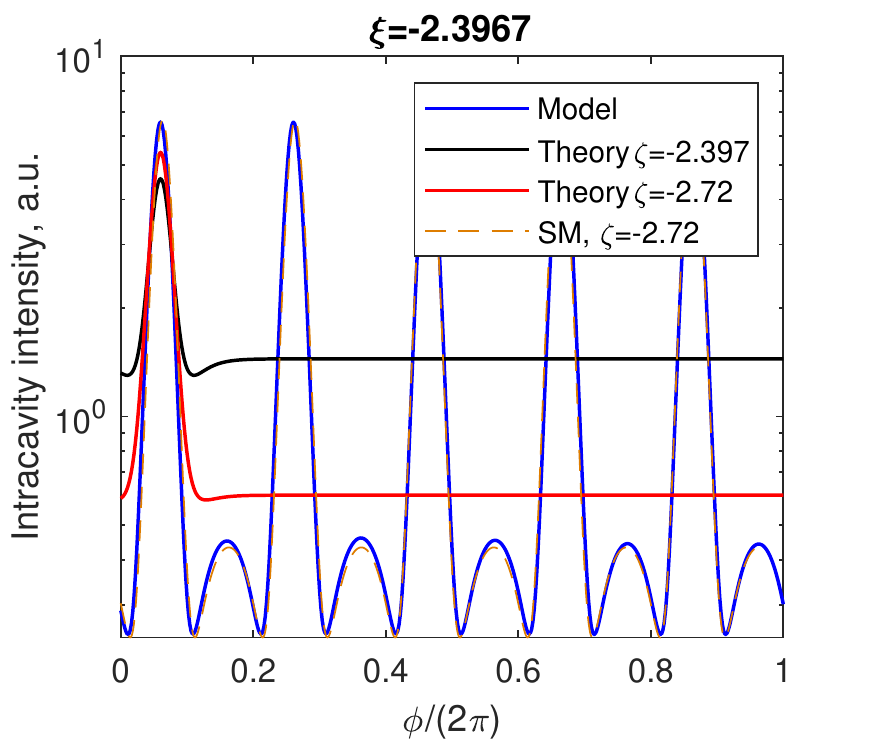}
\caption{Left: The field forward wave $A^+$ evolution inside the cavity with the decrease of the laser cavity detuning $\xi$ at $\beta=0.1$, $\omega_0t_s=3\pi/4$, $f=1.8191$. The red dash-dotted curve is classical pump detuning limit $\zeta=\pi^2f^2/8$  for the soliton existence.
Right: The azimuthal field profile at the detuning $\xi\approx-2.4$ (blue) together with the theoretical profile of typical secant-square soliton \cite{herr2014temporal} at effective detunings $\zeta=-2.72$ (red) and $\zeta=-2.4$ (black) and numerically calculated profile for the simple non-locked model at $\zeta=-2.72$ (orange dashed).
}
\label{fig:f2soli}
\end{figure}
\begin{figure}[ht]
\centering
\includegraphics[width=0.48\linewidth]{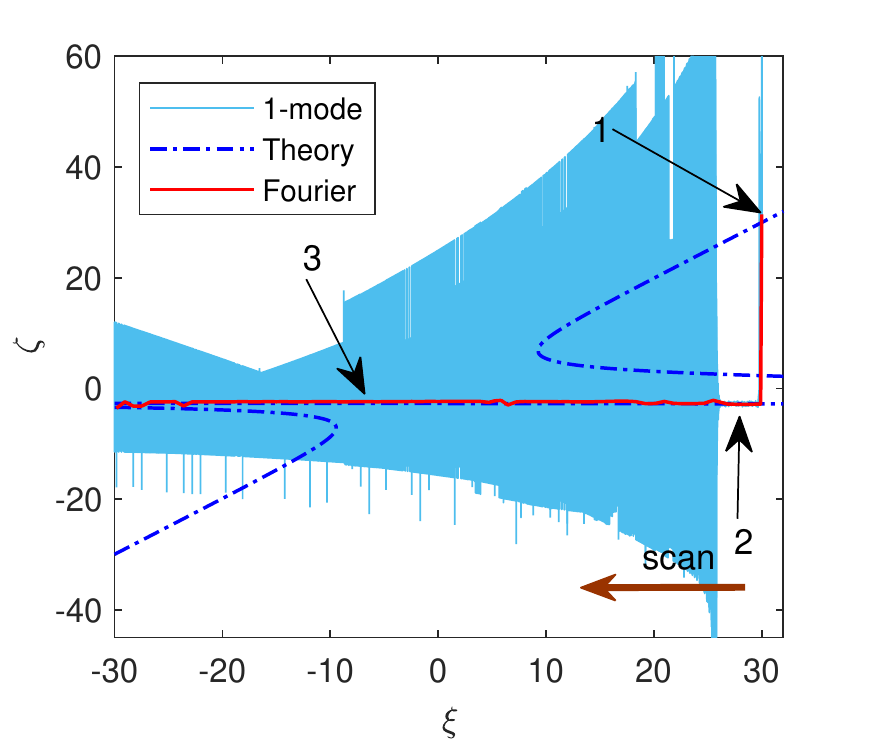}
\includegraphics[width=0.48\linewidth]{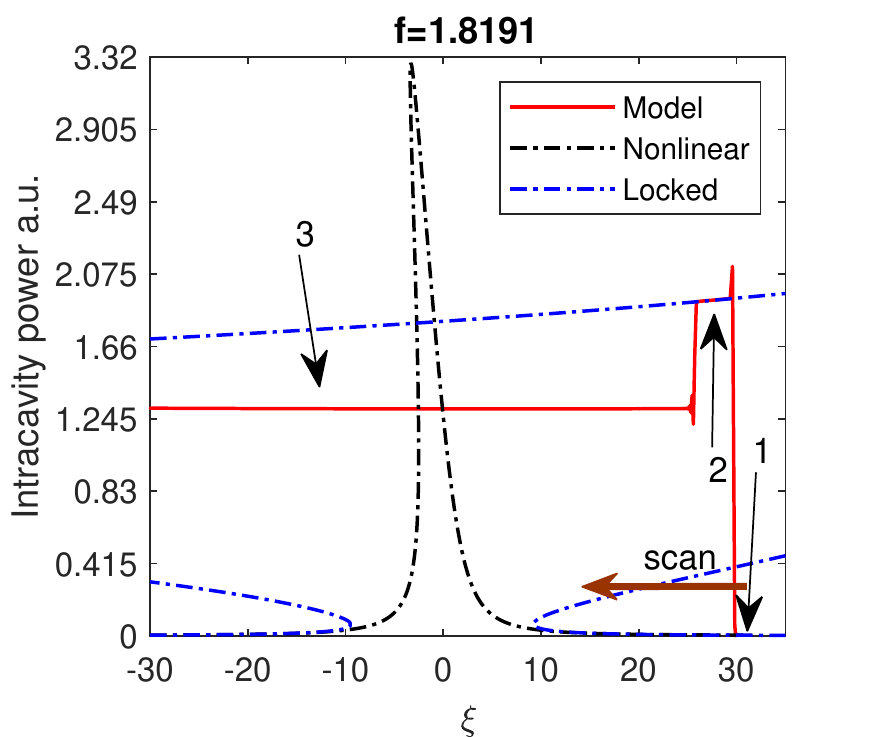}
\caption{Left: Theoretical tuning curve (blue dash-dotted) together with the single-mode (light-blue) and Fourier (red) estimations of the generation detuning $\zeta$ at $\beta=0.1$, $\omega_0t_s=3\pi/4$, $f=1.8191$. Right: Resonance curve (intracavity power vs. laser cavity detuning) for the same process. Red line -- numerical results for the developed model, blue dash-dotted line -- locked nonlinear resonance, black dash-dotted line -- unlocked nonlinear resonance. 1 -- spontaneous locking, 2 -- locked CW evolution, 3 -- locked solitons.}
\label{fig:f2tune}
\end{figure}
First, we consider the ``zero-phase case'' $\omega_0t_s=3\pi/4$, which was shown to be the best regime in terms of the laser stabilization and locking band for $\beta<1$\cite{PhysRevApplied.14.014036}. Figure \ref{fig:f2soli} shows the evolution of the intensity inside the microresonator upon the laser cavity detuning $\xi$ scan for low input power. We see the  generation of the soliton crystal \cite{Karpov2019}, representing temporally-ordered ensemble of soliton pulses, in the locked state. The solitons propagate without distortions while the laser cavity detuning $\xi$ is {scanned} in a wide range far further than the classical soliton detuning limit $\pi^2f^2/8$ [see red dash-dotted line in Fig. \ref{fig:f2soli}], because the effective detuning $\zeta$ stays practically constant. This indicates that the laser operates in the locked state. In the right panel in Fig. \ref{fig:f2soli} the form of the standard secant-square solitonic pulse \cite{herr2014temporal} is compared with one of the solitons in the crystal. It can be seen that the form of the pulse is close to the analytical one for the correct locked detuning $\zeta=-2.72$. We also performed modeling of the unlocked system with backscattering and found the 100\% coincidence of the obtained waveforms [see orange dashed line in the right panel of Fig. \ref{fig:f2soli}].

Fig. \ref{fig:f2tune} shows the tuning curve $\zeta(\xi)$ and the resonance curve $P_a(\xi)=\sum|A^+_\mu(\xi)|$, corresponding to this regime. In the left panel in Fig. \ref{fig:f2tune} we can see the theoretical prediction of the nonlinear tuning curve\cite{Voloshin2020,refId0} [blue dash-dotted line] with two types of the {instantaneous} effective detuning estimations [single-mode estimation $\zeta_1$ -- light blue and Fourier estimation $\zeta_F$ -- red]. Note, that the locked (horizontal) parts of the red and {blue dash-dotted} curves are below the bistability detuning \cite{Voloshin2020,Chembo2014}
\begin{align}
\label{bistability}
\zeta_{\rm bs}=&-\left(\frac{f}{2}\right)^{2/3}-\sqrt{4\left(\frac{f}{2}\right)^{4/3}-1},
\end{align}
which is $\zeta_{\rm bs}=-2.52$ in this case. First, both detuning estimations follow the free-running laser tuning dependence [$\xi=\zeta$ part of the theoretical curve], but then the spontaneous locking happens before the turning point and the effective detuning jumps to the nearly-horizontal ``locked'' part of the theoretical curve [see region 1 in Fig. \ref{fig:f2tune}]. Then locked evolution happens without nonlinear generation [see region 2 in Fig. \ref{fig:f2tune}] and all three curves coincide. The CW-solution [continuous wave, e.g. monochromatic wave with amplitude independent on the azimuthal coordinates] can be seen in Fig. \ref{fig:f2soli} at that moment. Finally, the simple single-mode estimation [light blue curve] starts oscillating at the point, where the nonlinear generation begins and solitons are formed [compare region 3 in Fig. \ref{fig:f2tune} and Fig. \ref{fig:f2soli}].
The trace of the intracavity power [Fig. \ref{fig:f2tune}, right panel], that has direct correspondence to the experimental LI-traces [light power vs. diode current], also exhibits characteristic features referred to the solitonic generation. The clear solitonic step\cite{herr2014temporal} can be seen [region 3 in Fig. \ref{fig:f2tune}] on the nearly rectangular shape of the self-injection-locked resonance [see the region 2 and the blue dash-dotted line in the right panel of Fig. \ref{fig:f2tune}]. The theoretical resonance curve can be obtained as the CW-solution of the full equations \eqref{Model_carr}-\eqref{Model_back} \cite{Voloshin2020}.

\begin{figure}[ht]
\centering
\includegraphics[width=0.48\linewidth]{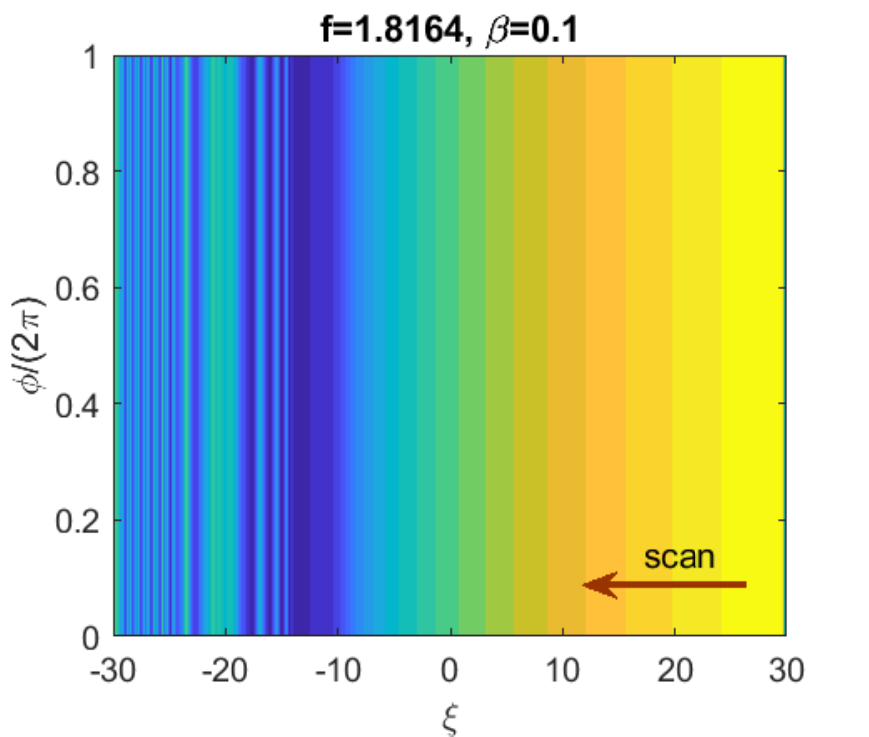}
\includegraphics[width=0.48\linewidth]{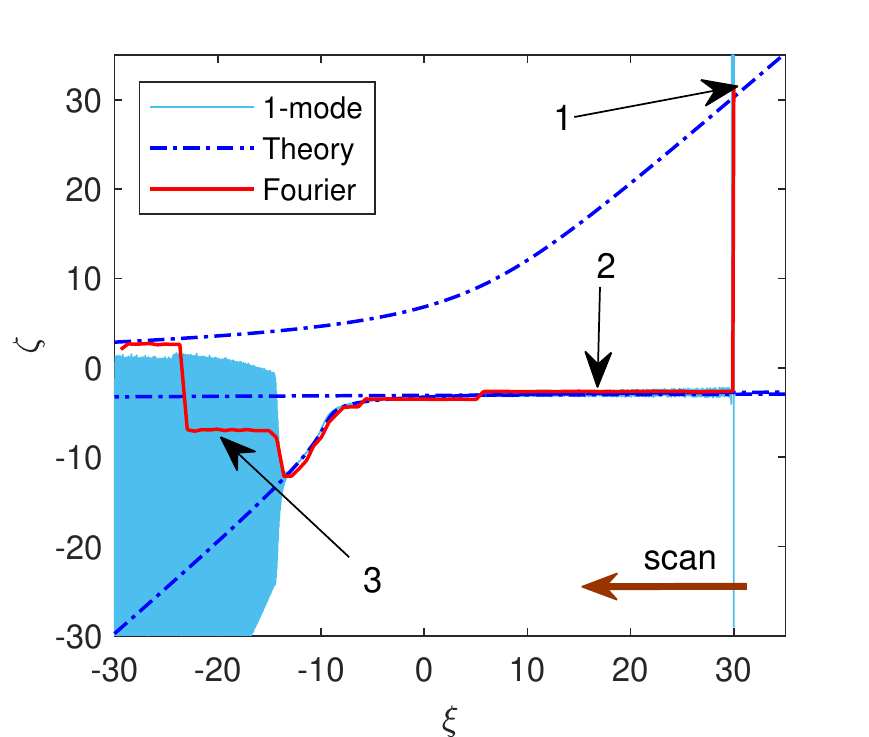}
\caption{Left: The field evolution of the forward wave $A^+$ inside the cavity with the laser cavity detuning $\xi$ decrease at $\beta=0.1$, $\omega_0t_s=0.6\pi/4$, $f=1.82$ for comparison with Fig. \ref{fig:f2soli} (left). Right: Theoretical tuning curve (blue dash-dotted) together with the single-mode (light blue) and Fourier (red) estimations of the laser generation detuning for comparison with the left panel in Fig. \ref{fig:f2tune}. 1 -- spontaneous locking, 2 -- locked CW evolution, 3 -- tuning curve instability.}
\label{fig:compare}
\end{figure}

\begin{figure}[ht]
\centering
\includegraphics[width=0.48\linewidth]{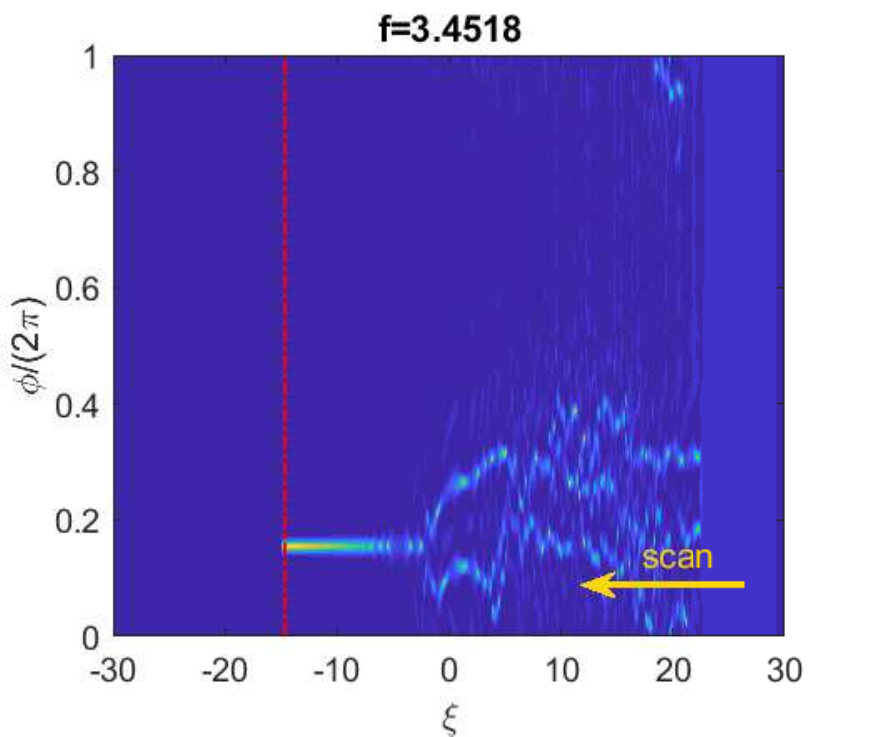}
\includegraphics[width=0.48\linewidth]{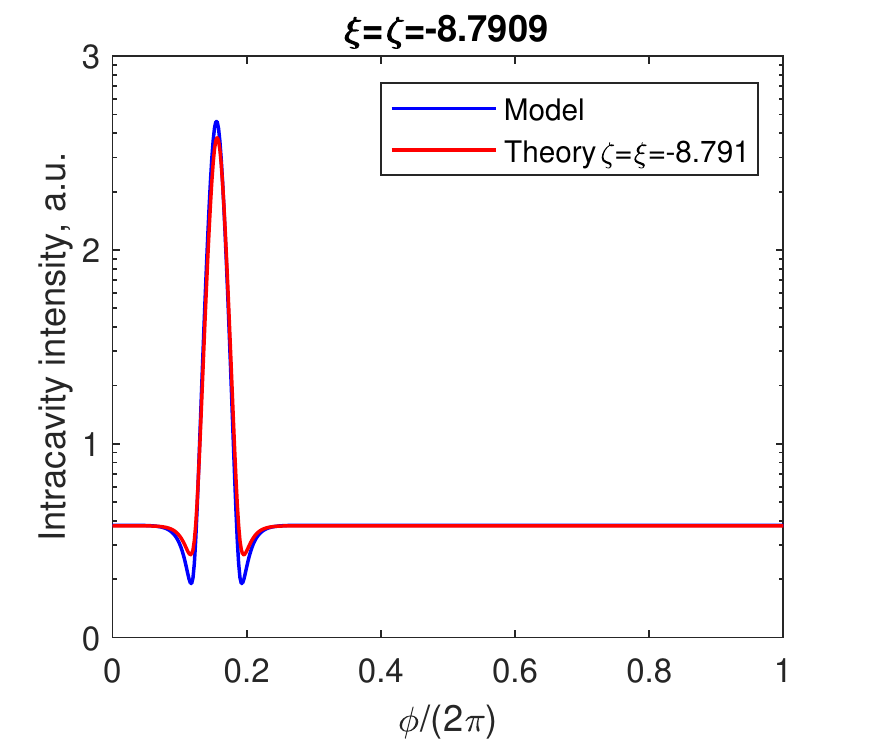}
\caption{Left: The field evolution inside the cavity forward wave $A^+$ with the laser cavity detuning $\xi$ decrease at $\beta=0.037$, $\omega_0t_s=0.6\pi/4$, $f=3.4518$. The vertical red dash-dotted line is maximum soliton detuning $\pi^2f^2/8$.
Right: The azimuthal field profile at the detuning $\xi\approx-8.8$ together with the theoretical profile of typical secant-square soliton \cite{herr2014temporal} at effective detuning $\zeta=-8.8$.}
\label{fig:f4soli}
\end{figure}
\begin{figure}[hb]
\centering
\includegraphics[width=0.48\linewidth]{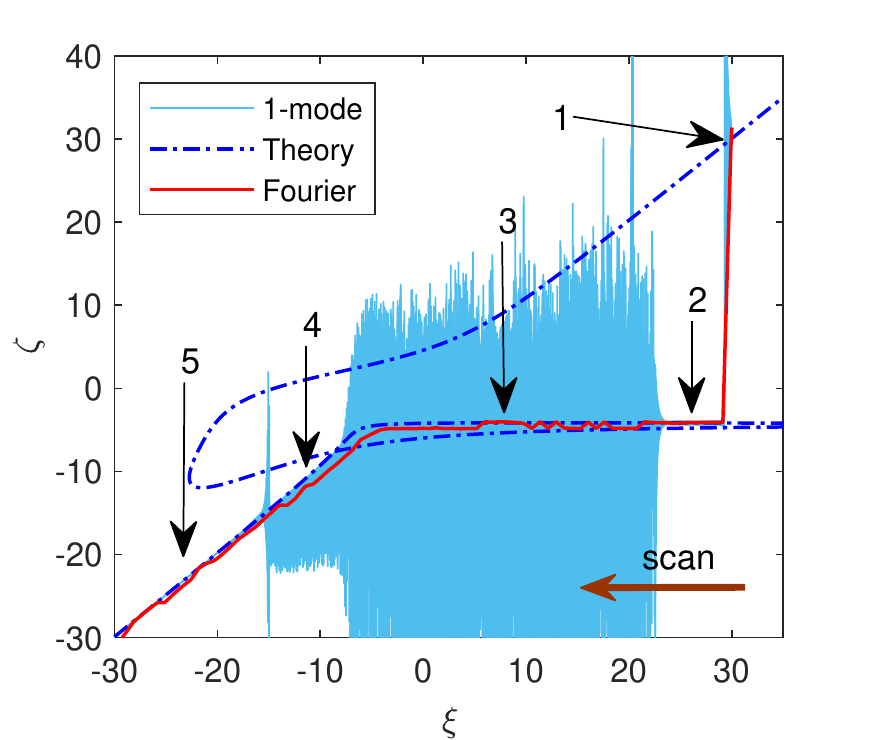}
\includegraphics[width=0.48\linewidth]{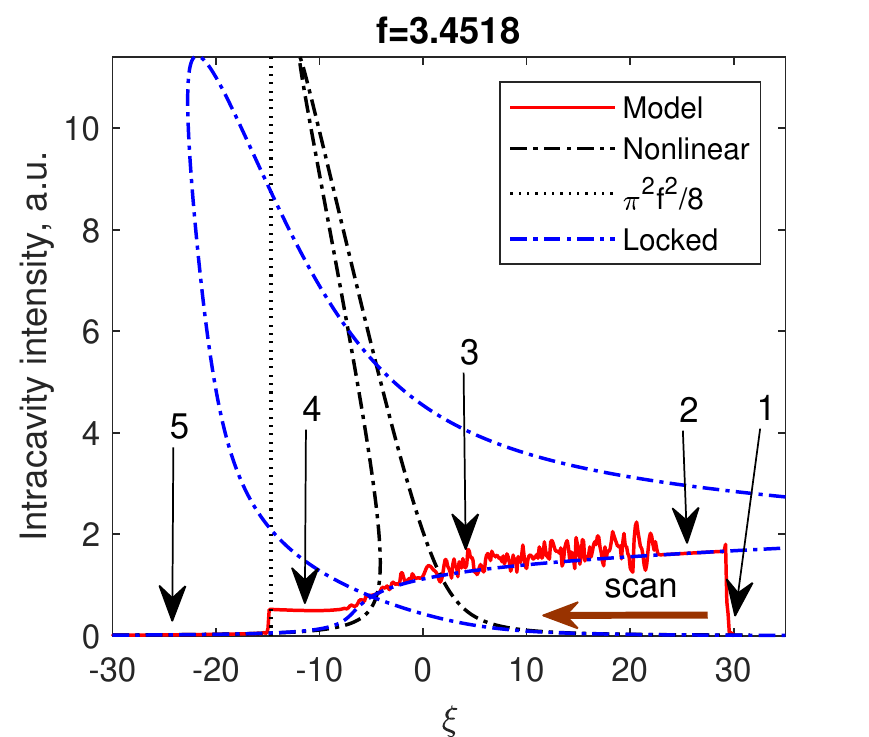}
\caption{Left: Theoretical tuning curve (blue dash-dotted) together with single-mode (light blue) and Fourier (red) estimations of the laser generation detuning at $\beta=0.037$, $\omega_0t_s=0.6\pi/4$, $f=3.4518$. Right: Resonance curve (intracavity power vs. laser cavity detuning) for the same process. Red line -- numerical results for the developed model, blue dash-dotted line -- locked resonance, black dash-dotted line -- unlocked nonlinear resonance, vertical dotted line -- maximum soliton detuning $\pi^2f^2/8$. 1 -- spontaneous locking, 2 -- locked CW evolution, 3 -- locked chaos, 4 -- unlocked soliton, 5 -- unlocked CW-evolution.}
\label{fig:f4tune}
\end{figure}
It is well-known that the locking phase $\omega_0t_s$ is a crucial parameter for the SIL effect [see \cite{Kondratiev:17}]. We also show that regimes of the SIL solitonic generation also depend on this parameter. The left panel in Fig. \ref{fig:compare} shows the intracavity field evolution and tuning curve for the ``half-wave'' phase ($\omega_0t_s=0.6\pi/4$) and other parameters the same as for the Fig. \ref{fig:f2soli}. It can be seen that only CW-solution exists [see left panel of Fig. \ref{fig:compare}], though the effective detuning $\zeta$ stays around $-3$ [see right panel of Fig. \ref{fig:compare}, region 2]. At some moment after the locking is lost, some instability arises between the different branches of the tuning curve. There are several lines of similar magnitude in the full Fourier spectrum in the region 3 of the right panel of Fig. \ref{fig:compare}.
By choosing different pump amplitude and backscattering coefficient we find the simple (unlocked) soliton regime. Fig. \ref{fig:f4soli} shows a single-soliton state in such regime. We can see that the power of the soliton peak grows with the detuning until reaching the termination point $\pi^2f^2/8$, indicating no locking. 

Figure \ref{fig:f4tune} provides more explanation of this regime. Similar to the previous case, the generation detuning follows the theoretical curve [region 1 in Fig. \ref{fig:f4tune} -- free-running regime and spontaneous locking, 2 -- locked state] and the nonlinear generation is clearly seen in the single-mode estimation [region 3 in Fig. \ref{fig:f4tune}]. At some point, the locking [e.g. the weak dependence of the effective detuning on the laser cavity detuning] is lost and the detuning follows the dependence for the free-running laser $\zeta=\xi$ law [see red line in region 4 of Fig. \ref{fig:f4tune}], the single-mode detuning oscillation character changes [see blue line in region 4 of Fig. \ref{fig:f4tune}]. This corresponds to the soliton formation [see the left panel of Fig. \ref{fig:f4soli}]. Moreover, in the resonance curve [right panel of Fig. \ref{fig:f4tune}] the working point transits to the solitonic step after the self-injection-locked resonance [region 4 in the right panel of Fig. \ref{fig:f4tune}]. Then, finally the soliton decays to the CW-solution exactly at the maximum soliton detuning $\pi^2f^2/8$ [region 5 in Fig. \ref{fig:f4tune}].
\begin{figure}[ht]
\center
\includegraphics[width=1\textwidth]{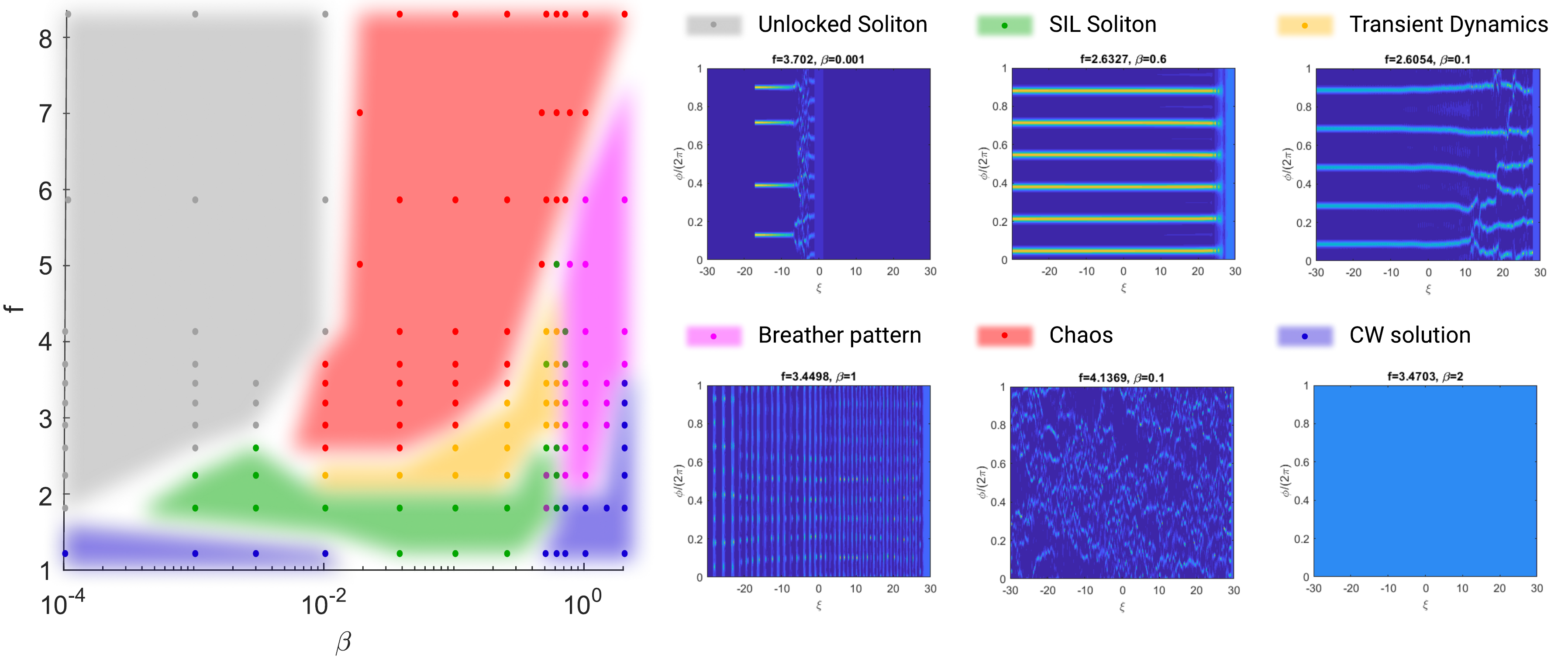}
\caption{The diagram of the regimes arising upon pump frequency sweep at anomalous GVD for different combinations of the pump amplitude $f$ and backscattering coefficient $\beta$.
%Pump-scattering diagram of the soliton dynamics (anomalous dispersion). 
The dots show the actually calculated parameters, the shaded regions -- extrapolation of the regime type. The white areas are either transient regions or too far from the calculated points to be certain. The legend is shown on the right and consists of panels with examples of characteristic regimes. For parameters see Table \ref{tab:Param}.}
\label{fig:SolitonDiagram}
\end{figure}

The dynamics of the solitons were studied for a wide range of the pump amplitude $f$ and backscattering coefficient $\beta$. The results are collected into a diagram shown in Fig. \ref{fig:SolitonDiagram}. 
For low $\beta$ no signs of SIL can be seen [gray region in Fig. \ref{fig:SolitonDiagram}] and "unlocked" soliton exhibits common dynamics.
Then, depending on the pump amplitude, a self-injection-locked solitons or chaotic regimes can be seen [green and red regions in Fig. \ref{fig:SolitonDiagram}]. {The solitons usually appear in the form of soliton crystals \cite{Karpov2019}, which probably happens due to the high intracvity power and low effective detuning.}
Between the SIL solitons and chaos a transient dynamics regime exists [orange region in Fig. \ref{fig:SolitonDiagram}]. In this regime solitons are seen, but they may exhibit breathing behaviour, drifts, spontaneous birth and decay. Such regime can be attributed to the transient chaos \cite{Karpov2019} and sometimes it transforms into solitonic regime with the decrease of the detuning (or growth of its absolute value).
For higher backscattering values the solution has the form of breathing patterns [magenta region in Fig. \ref{fig:SolitonDiagram}] or follows the CW-solution at small pump amplitudes.

It should be noted, that all the traces were got while sweeping the detuning. However, in several points we also performed a sweep stop analysis, allowing the system to continue evolving with fixed detuning. In this case we found that in the majority of cases the character of the evolution did not change, e.g. stable solitons remained stable, chaotic and breathing patterns remained chaotic and breathing patterns. However, in some cases transient dynamics behaviour [orange region in Fig. \ref{fig:SolitonDiagram}] led to stable solitons in the end.

Another important observation is that the the regions of locked solitons, transient dynamics and chaos [green, yellow and red areas in Fig. \ref{fig:SolitonDiagram}] correlate with the region, where the resulting effective detuning $\zeta$ actually lies bellow the bistability limit \eqref{bistability}, while the CW solution was got when $\zeta$ was above the threshold and breather patterns close to the boundary. 

\subsection{Normal GVD}
\begin{figure}[ht]
\centering
\includegraphics[width=0.48\linewidth]{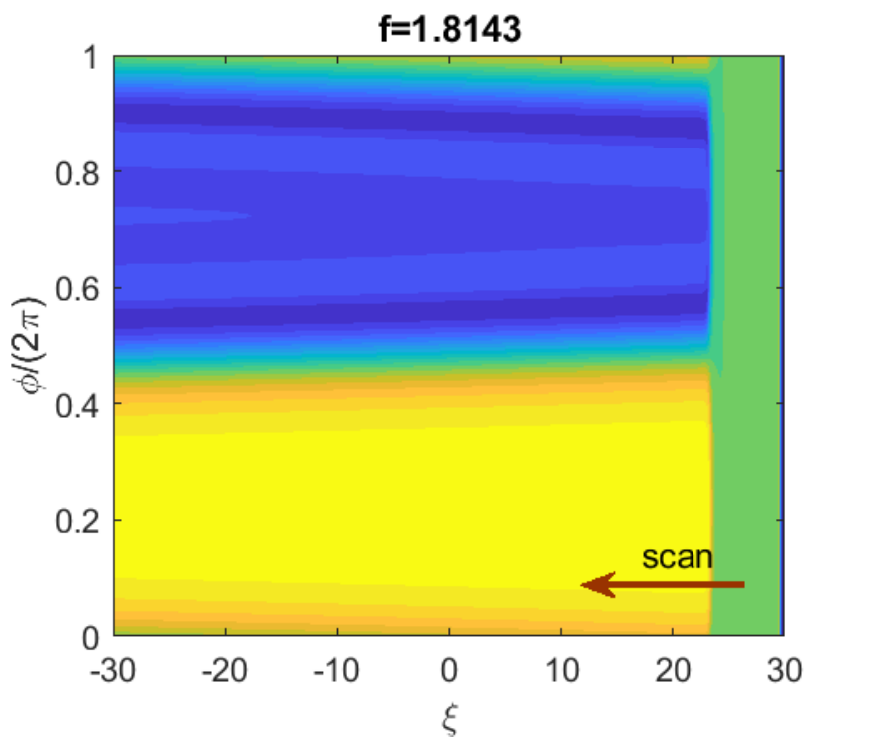}
\includegraphics[width=0.48\linewidth]{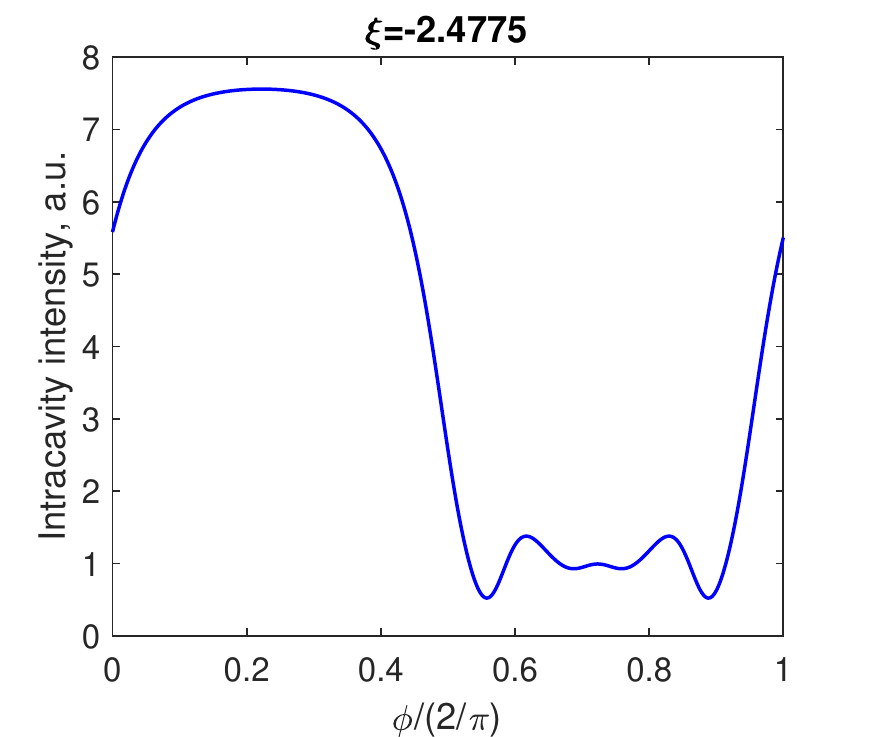}
\caption{Left: The forward wave $A^+$ field evolution inside the cavity with the laser cavity detuning $\xi$ decrease at $\beta=0.1$, $\omega_0t_s=3\pi/4$, $f=1.8143$.
Right: The azimuthal field profile at detuning $\xi=-2.4775$ (effective detuning $\zeta\approx-2.72$).}
\label{fig:normf2soli}
\end{figure}

\begin{figure}[ht]
\centering
\includegraphics[width=0.48\linewidth]{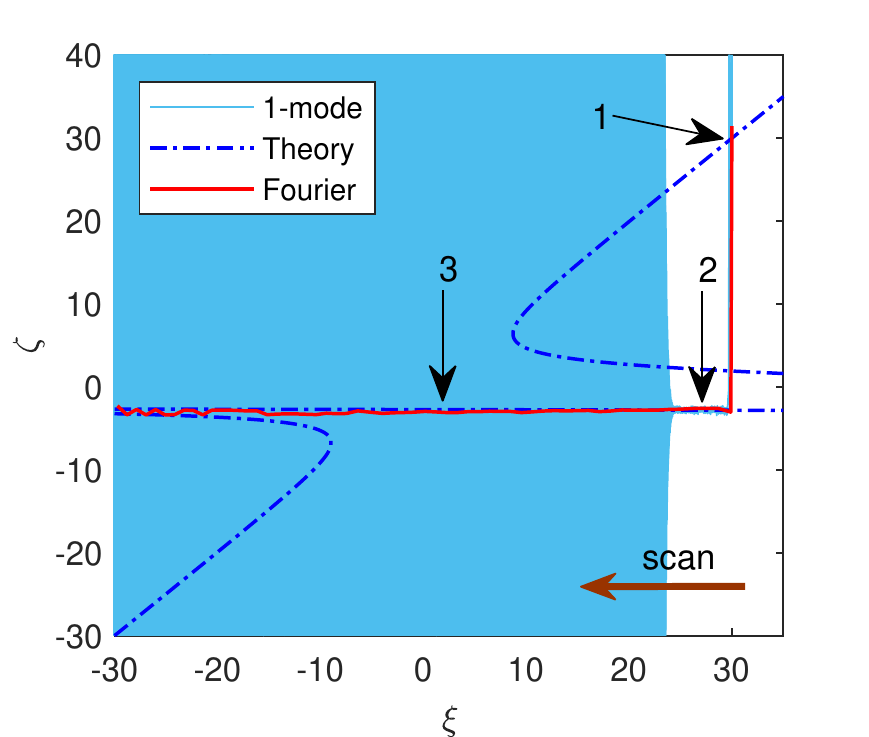}
\includegraphics[width=0.48\linewidth]{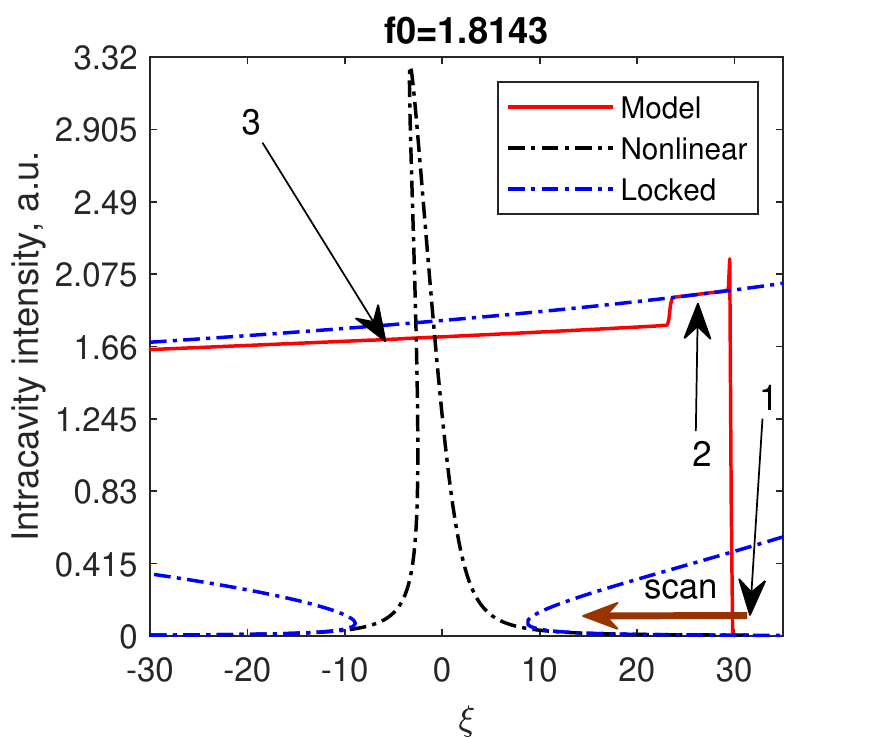}
\caption{Left: Theoretical tuning curve (blue dash-dotted) together with single-mode (light blue) and Fourier (red) estimations of the laser generation (effective) detuning at $\beta=0.1$, $\omega_0t_s=3\pi/4$, $f=1.8143$. Right: Resonance curve (intracavity power vs. laser cavity detuning) for the same process. Red line -- numerical results for the developed model, blue dash-dotted line -- locked resonance, black dash-dotted line -- unlocked nonlinear resonance. 1 -- spontaneous locking, 2 -- locked CW evolution, 3 -- locked platicon.}
\label{fig:normf2tune}
\end{figure}

It is well-known, that at normal GVD it is difficult to realize frequency comb generation while scanning the pump frequency if no special methods are used \cite{Lobanov2015,Lobanov2019}. When we switch off the backscattering [described by parameter $\beta$] or the back-action [described by $\tilde \kappa_{\rm Laser}$], we see only stable nonlinear resonance. However if those two are nonzero the solitonic pulse generation is observed in certain range of backscattering coefficient $\beta$ and pump amplitude $f$. Figures \ref{fig:normf2soli} and \ref{fig:normf2tune} show the evolution dynamics, exemplary platicon profile, tuning curve and resonance curves for the SIL platicon generation. The process of the platicon generation is very similar to that of the soliton. First, the spontaneous locking happens [see region 1 in Fig. \ref{fig:normf2tune}]. Then the system evolve in the locked CW state for some time [see region 2 in Fig. \ref{fig:normf2tune}] and finally the nonlinear generation happens and the platicon emerges  [see region 3 in Fig. \ref{fig:normf2tune}].
{The  generation of dark solitons or platicons by the free-running laser was found to be practically impossible due to the absence of the modulational instability (MI), so that the system remained in the cw solution during the scanning. In our earlier work we showed that the presence of the backward wave can induce MI \cite{Kondratiev2019}. In our opinion this fact in combination with nontrivial tuning curve in the SIL regime [see Fig. \ref{fig:normf2tune}, right] results in a complex nonlinear dynamics, which is very different from the dynamics of processes for a free-running laser, leading to the generation of platicons.}

\begin{figure}[ht]
\center
\includegraphics[width=0.9\textwidth]{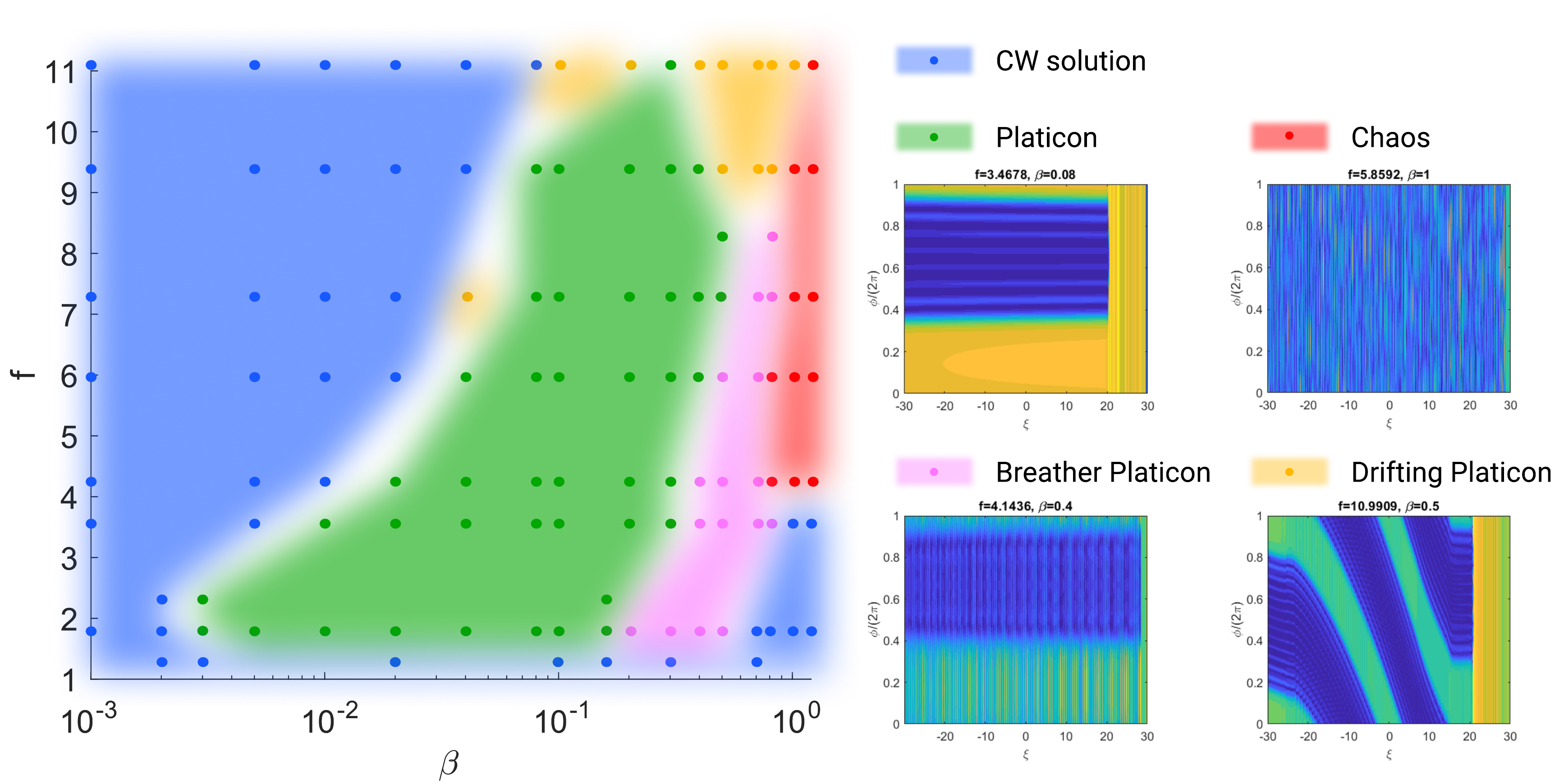}
\caption{The diagram of the regimes arising upon pump frequency sweep at normal GVD for different combinations of the pump amplitude $f$ and backscattering coefficient $\beta$.
%The pump-scattering diagram of the platicon dynamics (normal dispersion). 
The dots show the actually calculated parameters, the shaded regions -- extrapolation of the regime type. The white areas are either transient regions or too far from the calculated points to be certain. The legend is shown on the right and consists of panels with examples of characteristic regimes. For parameters see Table \ref{tab:Param}.}
\label{fig:PlaticonDiagram}
\end{figure}

\begin{figure}[ht]
\centering
\includegraphics[width=0.48\linewidth]{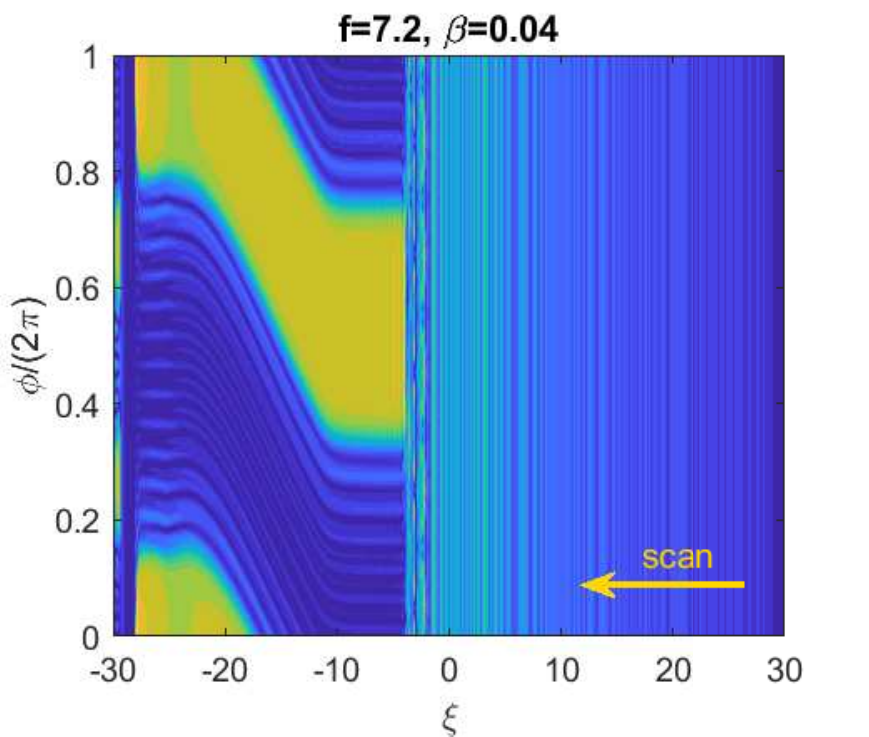}
\includegraphics[width=0.48\linewidth]{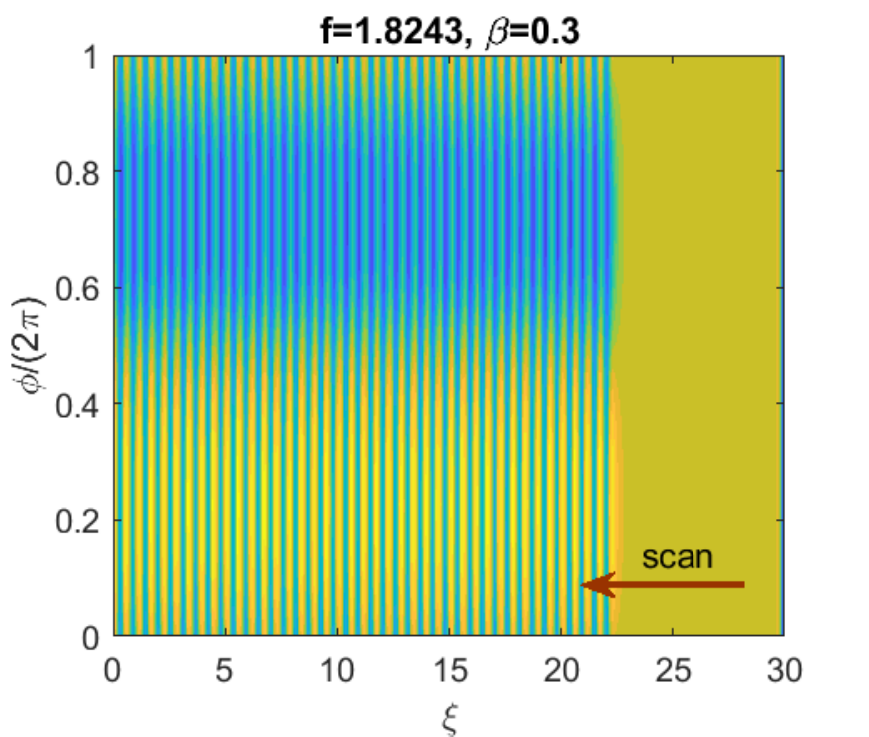}
\caption{Left: CW solution instabilities and platicon drift at $f=7.2$ and $\beta=0.04$.
Right: Breather-like platicon at $f=1.82$ and $\beta=0.3$. The locking phase is $\omega_0t_s=3\pi/4$.}
\label{fig:norm_high}
\end{figure}

The dynamics of the platicons were studied for ``zero-phase case'' $\omega_0t_s=3\pi/4$, wide range of pump amplitudes and backscattering coefficients. The results are collected into a diagram shown in Fig. \ref{fig:PlaticonDiagram}. 
We found that the minimal forward-backward wave coupling coefficient $\beta_{\rm min}$ enabling the nonlinear generation grows with the pump power. It ranges from $\beta_{\rm min}\approx0.005$ for $f=1.82$ to $\beta_{\rm min}\approx0.08$ at $f=10$. 
For higher backscattering, depending on the pump amplitude, a breathing and drifting regimes can be seen [magenta and orange regions in Fig. \ref{fig:PlaticonDiagram}]. At small pump amplitude values the platicon exhibit breather-like behaviour [see Fig. \ref{fig:norm_high}, right panel] and degrade to CW solution at higher values. We note that this pulsation of platicon power does not originate from the energy transfer between the forward and backward modes as it occurs in both waves practically synchronously.
Higher values of pump power introduce more instabilities in CW (continuous wave) solution before the generation and drift of the platicon [repetition rate change]. In some cases [see Fig. \ref{fig:norm_high}, left panel] this effect exists only in a certain range of detunings between the regions of stable propagation. Furthermore, sometimes breathing behaviour does not completely vanish while drifting. Note, that earlier it was reported that platicon drift is possible due to the influence of the third-order dispersion \cite{Lobanov2017}.
Finally, at high backscattering value only CW-solution and chaotic solutions exist.

We also performed a sweep-stop analysis, allowing the system to continue evolving with fixed detuning. In this case we also found that in the majority of cases the character of the evolution did not change.

{The platicons are famous for their pump-to-comb conversion efficiency. The self-injection locking (SIL) did not show any significant changes to this property. The single-soliton regime provides 5-6 times worse result then the platicon regime. However multisoliton regime can give more efficiency. Defining the efficiency as the ratio of the total power of the non-pumped modes to the input power $(P_a-|a^+_0|^2)/P_{in}=4\eta_0^2\frac{P_a-|a^+_0|^2}{f_e^2|a_l|^2}$ obtain, for example, for multisoliton regime in Fig. \ref{fig:f2soli}(right) 8.5\%, for single soliton Fig. \ref{fig:f4soli}(right) 1.4\% and for platicon in Fig. \ref{fig:normf2soli}(right) 7.3\%.}

\section{Conclusion}
Using original theoretical model we studied nonlinear processes in high-Q optical microresonators in the self-injection locking regime. Generation of the dissipative Kerr solitons (anomalous GVD) and platicons (normal GVD) was demonstrated numerically for the self-injection-locked pump. While the former was experimentally demonstrated in several works, the latter is not well studied. We revealed and identified different regimes of the generation of solitons and platicons for different combinations of the locking phase (laser-microresonator roundtrip time), backscattering coefficient and pump power.  Generation of both types of the considered solitonic pulses was shown to be possible in a certain range of the locking phase and become less stable at high pump powers. Generation of the dissipative Kerr solitons was not found to be very sensitive to the normalized backscattering $\beta$, while for the platicon generation this is a key parameter. The threshold value of the backscattering coefficient was found to grow with the pump power. Some nontrivial dynamics such as drift and breathing dynamics of the self-injection-locked platicons was revealed. Finally, we built diagrams of the expected solitonic pulse generation dynamics for a wide range of the pump and backscattering parameters. We hope that they will provide deep insight into the rich dynamics of the system and help in designing effective microresonator-based devices based on the SIL effect.

\section*{Funding}
Russian Science Foundation (17-12-01413-$\Pi$).

\section*{Acknowledgments}
VEL acknowledges the personal support from the Foundation for the Advancement of Theoretical Physics and Mathematics "BASIS". 

\section*{Disclosures}
The authors declare no conflicts of interest.

\bibliography{Thebibliography}

\begin{thebibliography}{10}
\newcommand{\enquote}[1]{``#1''}

\bibitem{Ohta1}
T.~{Ohta} and K.~{Murakami}, \enquote{Reducing negative resistance oscillator
  noise by self-injection,} {\protect\JournalTitle{Electron. Commun. Jpn.}}
  \textbf{51-B}, 80--82 (1968).

\bibitem{Chang2}
{Heng-Chia Chang}, \enquote{Stability analysis of self-injection-locked
  oscillators,} {\protect\JournalTitle{IEEE Transactions on Microwave Theory
  and Techniques}} \textbf{51}, 1989--1993 (2003).

\bibitem{magnetron1}
J.~J. {Choi} and G.~W. {Choi}, \enquote{Experimental observation of frequency
  locking and noise reduction in a self-injection-locked magnetron,}
  {\protect\JournalTitle{IEEE Transactions on Electron Devices}} \textbf{54},
  3430--3432 (2007).

\bibitem{magnetron2}
Y.~P. {Bliokh}, Y.~E. {Krasik}, and J.~{Felsteiner},
  \enquote{Self-injection-locked magnetron as an active ring resonator side
  coupled to a waveguide with a delayed feedback loop,}
  {\protect\JournalTitle{IEEE Transactions on Plasma Science}} \textbf{40},
  78--82 (2012).

\bibitem{gyrotron2}
M.~M. {Melnikova}, A.~G. {Rozhnev}, N.~M. {Ryskin}, A.~V. {Tyshkun}, M.~Y.
  {Glyavin}, and Y.~V. {Novozhilova}, \enquote{Frequency stabilization of a
  0.67-{THz} gyrotron by self-injection locking,} {\protect\JournalTitle{IEEE
  Trans. Electr. Dev.}} \textbf{63}, 1288--1293 (2016).

\bibitem{7119883}
L.~{Zhang}, A.~K. {Poddar}, U.~L. {Rohde}, and A.~S. {Daryoush}, \enquote{Phase
  noise reduction in {RF} oscillators utilizing self-injection locked and phase
  locked loop,} in \emph{2015 IEEE 15th Topical Meeting on Silicon Monolithic
  Integrated Circuits in {RF} Systems,}  (2015), pp. 86--88.

\bibitem{Patzak1983}
E.~{Patzak}, H.~{Olesen}, A.~{Sugimura}, S.~{Saito}, and T.~{Mukai},
  \enquote{Spectral linewidth reduction in semiconductor lasers by an external
  cavity with weak optical feedback,} {\protect\JournalTitle{Electronics
  Letters}} \textbf{19}, 938--940 (1983).

\bibitem{Agrawal1984}
G.~{Agrawal}, \enquote{Line narrowing in a single-mode injection laser due to
  external optical feedback,} {\protect\JournalTitle{IEEE Journal of Quantum
  Electronics}} \textbf{20}, 468--471 (1984).

\bibitem{Oraevsky2001}
A.~N. Oraevsky, A.~V. Yarovitsky, and V.~L. Velichansky, \enquote{Frequency
  stabilisation of a diode laser by a whispering-gallery mode,}
  {\protect\JournalTitle{Quantum Electronics}} \textbf{31}, 897--903 (2001).

\bibitem{Liang2010}
W.~Liang, V.~S. Ilchenko, A.~A. Savchenkov, A.~B. Matsko, D.~Seidel, and
  L.~Maleki, \enquote{Whispering-gallery-mode-resonator-based ultranarrow
  linewidth external-cavity semiconductor laser,} {\protect\JournalTitle{Opt.
  Lett.}} \textbf{35}, 2822--2824 (2010).

\bibitem{Liang2015}
W.~Liang, V.~S. Ilchenko, D.~Eliyahu, A.~A. Savchenkov, A.~B. Matsko,
  D.~Seidel, and L.~Maleki, \enquote{Ultralow noise miniature external cavity
  semiconductor laser,} {\protect\JournalTitle{Nature Communications}}
  \textbf{6}, 7371 (2015).

\bibitem{Kondratiev:17}
N.~M. Kondratiev, V.~E. Lobanov, A.~V. Cherenkov, A.~S. Voloshin, N.~G. Pavlov,
  S.~Koptyaev, and M.~L. Gorodetsky, \enquote{Self-injection locking of a laser
  diode to a high-{Q} {W}{G}{M} microresonator,} {\protect\JournalTitle{Opt.
  Express}} \textbf{25}, 28167--28178 (2017).

\bibitem{PhysRevApplied.14.014036}
R.~R. Galiev, N.~M. Kondratiev, V.~E. Lobanov, A.~B. Matsko, and I.~A. Bilenko,
  \enquote{Optimization of laser stabilization via self-injection locking to a
  whispering-gallery-mode microresonator,} {\protect\JournalTitle{Phys. Rev.
  Applied}} \textbf{14}, 014036 (2020).

\bibitem{jin2020hertzlinewidth}
W.~Jin, Q.-F. Yang, L.~Chang, B.~Shen, H.~Wang, M.~A. Leal, L.~Wu, A.~Feshali,
  M.~Paniccia, K.~J. Vahala, and J.~E. Bowers, \enquote{Hertz-linewidth
  semiconductor lasers using {CMOS}-ready ultra-high-{Q} microresonators,}
  (2020).

\bibitem{BRAGINSKY1989393}
V.~Braginsky, M.~Gorodetsky, and V.~Ilchenko, \enquote{Quality-factor and
  nonlinear properties of optical whispering-gallery modes,}
  {\protect\JournalTitle{Physics Letters A}} \textbf{137}, 393 -- 397 (1989).

\bibitem{1588878}
A.~B. {Matsko} and V.~S. {Ilchenko}, \enquote{Optical resonators with
  whispering-gallery modes-part {I}: basics,} {\protect\JournalTitle{IEEE
  Journal of Selected Topics in Quantum Electronics}} \textbf{12}, 3--14
  (2006).

\bibitem{Lin2017}
G.~Lin, A.~Coillet, and Y.~K. Chembo, \enquote{Nonlinear photonics with
  high-{Q} whispering-gallery-mode resonators,} {\protect\JournalTitle{Adv.
  Opt. Photon.}} \textbf{9}, 828--890 (2017).

\bibitem{herr2014temporal}
T.~Herr, V.~Brasch, J.~D. Jost, C.~Y. Wang, N.~M. Kondratiev, M.~L. Gorodetsky,
  and T.~J. Kippenberg, \enquote{Temporal solitons in optical microresonators,}
  {\protect\JournalTitle{Nat. Photon.}} \textbf{8}, 145--152 (2014).

\bibitem{Kippenbergeaan8083}
T.~J. Kippenberg, A.~L. Gaeta, M.~Lipson, and M.~L. Gorodetsky,
  \enquote{Dissipative {K}err solitons in optical microresonators,}
  {\protect\JournalTitle{Science}} \textbf{361} (2018).

\bibitem{Pavlov2018}
N.~G. Pavlov, S.~Koptyaev, G.~V. Lihachev, A.~S. Voloshin, A.~S. Gorodnitskiy,
  M.~V. Ryabko, S.~V. Polonsky, and M.~L. Gorodetsky, \enquote{Narrow-linewidth
  lasing and soliton {K}err microcombs with ordinary laser diodes,}
  {\protect\JournalTitle{Nature Photonics}} \textbf{12}, 694--698 (2018).

\bibitem{Raja2019}
A.~S. Raja, A.~S. Voloshin, H.~Guo, S.~E. Agafonova, J.~Liu, A.~S.
  Gorodnitskiy, M.~Karpov, N.~G. Pavlov, E.~Lucas, R.~R. Galiev, A.~E.
  Shitikov, J.~D. Jost, M.~L. Gorodetsky, and T.~J. Kippenberg,
  \enquote{Electrically pumped photonic integrated soliton microcomb,}
  {\protect\JournalTitle{Nature Communications}} \textbf{10}, 680 (2019).

\bibitem{Shen2020}
B.~Shen, L.~Chang, J.~Liu, H.~Wang, Q.-F. Yang, C.~Xiang, R.~N. Wang, J.~He,
  T.~Liu, W.~Xie, J.~Guo, D.~Kinghorn, L.~Wu, Q.-X. Ji, T.~J. Kippenberg,
  K.~Vahala, and J.~E. Bowers, \enquote{Integrated turnkey soliton microcombs,}
  {\protect\JournalTitle{Nature}} \textbf{582}, 365--369 (2020).

\bibitem{Voloshin2020}
A.~S. Voloshin, J.~Liu, N.~M. Kondratiev, G.~V. Lihachev, V.~E. Lobanov,
  W.~Weng, T.~J. Kippenberg, and I.~A. Bilenko, \enquote{Dynamics of soliton
  self-injection locking in a photonic chip-based microresonator,}
  {\protect\JournalTitle{arXiv}} \textbf{1912.11303} (2020).

\bibitem{Suh600}
M.-G. Suh, Q.-F. Yang, K.~Y. Yang, X.~Yi, and K.~J. Vahala,
  \enquote{Microresonator soliton dual-comb spectroscopy,}
  {\protect\JournalTitle{Science}} \textbf{354}, 600--603 (2016).

\bibitem{9195660}
A.~{Lukashchuk}, J.~{Riemensberger}, M.~{Karpov}, E.~{Lucas}, W.~{Weng},
  J.~{Liu}, and T.~{Kippenberg}, \enquote{Microresonator soliton based
  massively parallel coherent lidar,} in \emph{2020 IEEE Research and
  Applications of Photonics in Defense Conference (RAPID),}  (2020), pp. 1--3.

\bibitem{Marin-Palomo2017}
P.~Marin-Palomo, J.~N. Kemal, M.~Karpov, A.~Kordts, J.~Pfeifle, M.~H.~P.
  Pfeiffer, P.~Trocha, S.~Wolf, V.~Brasch, M.~H. Anderson, R.~Rosenberger,
  K.~Vijayan, W.~Freude, T.~J. Kippenberg, and C.~Koos,
  \enquote{Microresonator-based solitons for massively parallel coherent
  optical communications,} {\protect\JournalTitle{Nature}} \textbf{546},
  274--279 (2017).

\bibitem{Suh2019}
M.-G. Suh, X.~Yi, Y.-H. Lai, S.~Leifer, I.~S. Grudinin, G.~Vasisht, E.~C.
  Martin, M.~P. Fitzgerald, G.~Doppmann, J.~Wang, D.~Mawet, S.~B. Papp, S.~A.
  Diddams, C.~Beichman, and K.~Vahala, \enquote{Searching for exoplanets using
  a microresonator astrocomb,} {\protect\JournalTitle{Nature Photonics}}
  \textbf{13}, 25--30 (2019).

\bibitem{Lihachev2020cleo}
J.~Liu, G.~Lihachev, L.~Chang, J.~He, R.~N. Wang, J.~Guo, A.~Raja, E.~Lucas,
  N.~Pavlov, J.~Jost, D.~Kinghorn, J.~Bowers, and T.~Kippenberg,
  \enquote{{Laser Self-Injection Locked Frequency Combs in a Normal GVD
  Integrated Microresonator},} in \emph{CLEO, STh1O.3,}  (2020).

\bibitem{Lobanov_2015}
V.~E. Lobanov, G.~Lihachev, and M.~L. Gorodetsky, \enquote{Generation of
  platicons and frequency combs in optical microresonators with normal {GVD} by
  modulated pump,} {\protect\JournalTitle{{EPL} (Europhysics Letters)}}
  \textbf{112}, 54008 (2015).

\bibitem{Liu:17}
H.~Liu, S.-W. Huang, J.~Yang, M.~Yu, D.-L. Kwong, and C.~W. Wong,
  \enquote{Bright square pulse generation by pump modulation in a normal gvd
  microresonator,} in \emph{Conference on Lasers and Electro-Optics,}  (Optical
  Society of America, 2017), p. FTu3D.3.

\bibitem{Lobanov2019}
V.~E. Lobanov, N.~M. Kondratiev, A.~E. Shitikov, R.~R. Galiev, and I.~A.
  Bilenko, \enquote{Generation and dynamics of solitonic pulses due to pump
  amplitude modulation at normal group-velocity dispersion,}
  {\protect\JournalTitle{Phys. Rev. A}} \textbf{100}, 013807 (2019).

\bibitem{Liu:14}
Y.~Liu, Y.~Xuan, X.~Xue, P.-H. Wang, S.~Chen, A.~J. Metcalf, J.~Wang, D.~E.
  Leaird, M.~Qi, and A.~M. Weiner, \enquote{Investigation of mode coupling in
  normal-dispersion silicon nitride microresonators for {K}err frequency comb
  generation,} {\protect\JournalTitle{Optica}} \textbf{1}, 137--144 (2014).

\bibitem{Xue2015}
X.~Xue, Y.~Xuan, Y.~Liu, P.-H. Wang, S.~Chen, J.~Wang, D.~E. Leaird, M.~Qi, and
  A.~M. Weiner, \enquote{Mode-locked dark pulse {K}err combs in
  normal-dispersion microresonators,} {\protect\JournalTitle{Nature Photonics}}
  \textbf{9}, 594--600 (2015).

\bibitem{Lobanov2015}
V.~Lobanov, G.~Lihachev, T.~J. Kippenberg, and M.~Gorodetsky,
  \enquote{Frequency combs and platicons in optical microresonators with normal
  {GVD},} {\protect\JournalTitle{Opt. Express}} \textbf{23}, 7713--7721 (2015).

\bibitem{Lobanov2017}
V.~E. Lobanov, A.~V. Cherenkov, A.~E. Shitikov, I.~A. Bilenko, and M.~L.
  Gorodetsky, \enquote{Dynamics of platicons due to third-order dispersion,}
  {\protect\JournalTitle{The European Physical Journal D}} \textbf{71}, 185
  (2017).

\bibitem{Kim:s}
B.~Y. Kim, Y.~Okawachi, J.~K. Jang, M.~Yu, X.~Ji, Y.~Zhao, C.~Joshi, M.~Lipson,
  and A.~L. Gaeta, \enquote{Turn-key, high-efficiency {K}err comb source,}
  {\protect\JournalTitle{Opt. Lett.}} \textbf{44}, 4475--4478 (2019).

\bibitem{doi:10.1002/lpor.201500107}
X.~Xue, Y.~Xuan, P.-H. Wang, Y.~Liu, D.~E. Leaird, M.~Qi, and A.~M. Weiner,
  \enquote{Normal-dispersion microcombs enabled by controllable mode
  interactions,} {\protect\JournalTitle{Laser \& Photonics Reviews}}
  \textbf{9}, L23--L28 (2015).

\bibitem{doi:10.1002/lpor.201600276}
X.~Xue, P.-H. Wang, Y.~Xuan, M.~Qi, and A.~M. Weiner, \enquote{Microresonator
  {K}err frequency combs with high conversion efficiency,}
  {\protect\JournalTitle{Laser \& Photonics Reviews}} \textbf{11}, 1600276
  (2017).

\bibitem{Jang:s}
J.~K. Jang, Y.~Okawachi, M.~Yu, K.~Luke, X.~Ji, M.~Lipson, and A.~L. Gaeta,
  \enquote{Dynamics of mode-coupling-induced microresonator frequency combs in
  normal dispersion,} {\protect\JournalTitle{Opt. Express}} \textbf{24},
  28794--28803 (2016).

\bibitem{Kondratiev2019}
N.~M. Kondratiev and V.~E. Lobanov, \enquote{Modulational instability and
  frequency combs in whispering-gallery-mode microresonators with
  backscattering,} {\protect\JournalTitle{Phys. Rev. A}} \textbf{101}, 013816
  (2020).

\bibitem{PhysRevA.82.033801}
Y.~K. Chembo and N.~Yu, \enquote{Modal expansion approach to
  optical-frequency-comb generation with monolithic whispering-gallery-mode
  resonators,} {\protect\JournalTitle{Phys. Rev. A}} \textbf{82}, 033801
  (2010).

\bibitem{Bao2019}
H.~Bao, A.~Cooper, M.~Rowley, L.~Di~Lauro, J.~S. Totero~Gongora, S.~T. Chu,
  B.~E. Little, G.-L. Oppo, R.~Morandotti, D.~J. Moss, B.~Wetzel, M.~Peccianti,
  and A.~Pasquazi, \enquote{Laser cavity-soliton microcombs,}
  {\protect\JournalTitle{Nature Photonics}} \textbf{13}, 384--389 (2019).

\bibitem{Gorodetsky:00}
M.~L. Gorodetsky, A.~D. Pryamikov, and V.~S. Ilchenko, \enquote{Rayleigh
  scattering in high-{Q} microspheres,} {\protect\JournalTitle{J. Opt. Soc. Am.
  B}} \textbf{17}, 1051--1057 (2000).

\bibitem{Karpov2019}
M.~Karpov, M.~H.~P. Pfeiffer, H.~Guo, W.~Weng, J.~Liu, and T.~J. Kippenberg,
  \enquote{Dynamics of soliton crystals in optical microresonators,}
  {\protect\JournalTitle{Nature Physics}} \textbf{15}, 1071--1077 (2019).

\bibitem{PhysRevA.87.053852}
Y.~K. Chembo and C.~R. Menyuk, \enquote{Spatiotemporal lugiato-lefever
  formalism for kerr-comb generation in whispering-gallery-mode resonators,}
  {\protect\JournalTitle{Phys. Rev. A}} \textbf{87}, 053852 (2013).

\bibitem{BOGACKI1989321}
P.~Bogacki and L.~Shampine, \enquote{A 3(2) pair of {R}unge - {K}utta
  formulas,} {\protect\JournalTitle{Applied Mathematics Letters}} \textbf{2},
  321 -- 325 (1989).

\bibitem{Shampine}
L.~F. Shampine and M.~W. Reichelt, \enquote{The matlab ode suite,}
  {\protect\JournalTitle{SIAM Journal on Scientific Computing}} \textbf{18},
  1--22 (1997).

\bibitem{refId0}
N.~Kondratiev, A.~Gorodnitskiy, and V.~Lobanov, \enquote{Influence of the
  microresonator nonlinearity on the self-injection locking effect,}
  {\protect\JournalTitle{EPJ Web Conf.}} \textbf{220}, 02006 (2019).

\bibitem{Chembo2014}
C.~Godey, I.~V. Balakireva, A.~Coillet, and Y.~K. Chembo, \enquote{Stability
  analysis of the spatiotemporal {L}ugiato-{L}efever model for {K}err optical
  frequency combs in the anomalous and normal dispersion regimes,}
  {\protect\JournalTitle{Phys. Rev. A}} \textbf{89}, 063814 (2014).

\end{thebibliography}

\end{document}